\pdfoutput=1

\documentclass[useAMS,usenatbib]{mn2e}

\usepackage{times}
\usepackage{lettrine}
\usepackage{epsfig}
\usepackage{graphicx}
\usepackage{color}
\usepackage{array}
\usepackage{pstricks}
\usepackage{titlesec}
\usepackage{makeidx}
\usepackage{amsmath}
\usepackage{bm}
\usepackage{mathrsfs}
\usepackage{amssymb}
\usepackage{makeidx}
\usepackage{supertabular}
\usepackage{fancyhdr}
\usepackage{url}
\usepackage{natbib}
\usepackage[a4paper,pdftex,verbose,margin=2cm,twoside]{geometry}
\title[Large-scale magnetic structures from buoyancy instabilities]{How can large-scale twisted magnetic structures naturally emerge from buoyancy instabilities?}
\author[B. Favier \textit{et al.}]{B. Favier$^{1}$\thanks{Email address: b.favier@damtp.cam.ac.uk}, L. Jouve$^{2,3}$, W. Edmunds$^1$, L. J. Silvers$^4$ and M. R. E. Proctor$^1$\\
$^{1}$Department of Applied Mathematics and Theoretical Physics, University of Cambridge, \\
\ Centre for Mathematical Sciences, Wilberforce Road, Cambridge CB3 0WA, UK\\
$^{2}$Universit\'e de Toulouse, UPS-OMP, Institut de Recherche en Astrophysique et Plan\'etologie, Toulouse, France\\
$^{3}$CNRS, Institut de Recherche en Astrophysique et Plan\'etologie, 14 Avenue \'Edouard Belin, 31400 Toulouse, France\\
$^{4}$Centre for Mathematical Science, City University London, Northampton Square, London, EC1V 0HB, UK}

\begin{document}

\date{\today}

\pagerange{\pageref{firstpage}--\pageref{lastpage}} \pubyear{2012}

\maketitle

\label{firstpage}

\begin{abstract}
We consider the three-dimensional instability of a layer of horizontal magnetic field in a polytropic atmosphere where, contrary to previous studies, the field lines in the initial state are not unidirectional.
We show that if the twist is initially concentrated inside the unstable layer, the modifications of the instability reported by several authors (see e.g. \citet{cattaneo1990}) are only observed when the calculation is restricted to two dimensions.
In three dimensions, the usual interchange instability occurs, in the direction fixed by the field lines at the interface between the layer and the field-free region.
We therefore introduce a new configuration: the instability now develops in a weakly magnetised atmosphere where the direction of the field can vary with respect to the direction of the strong unstable field below, the twist being now concentrated at the upper interface.
Both linear stability analysis and non-linear direct numerical simulations are used to study this configuration.
We show that from the small-scale interchange instability, large-scale twisted coherent magnetic structures are spontaneously formed, with possible implications to the formation of active regions from a deep-seated solar magnetic field.
\end{abstract}


%
%
\section{Introduction}

Active regions at the surface of the Sun are believed to be the visible manifestation of deep-seated intense magnetic fields.
The generation of these predominantly toroidal fields is associated with the differential rotation in the tachocline.
However, the transport of these strong fields from the lower part of the convective zone up to the photosphere remains one of the major unknowns of the solar cycle.
It is clear that super-equipartition fields are buoyant and could therefore rise up to the surface, but the observed size, coherence and twist of the magnetic structures forming active regions remain largely
unexplained.

In order to obtain some understanding of the transport of magnetic structures by buoyancy, highly idealised models have been considered.
An early model \citep{parker1955} considers the buoyancy properties of the magnetic field when it is concentrated in slender pressure-confined flux tubes.
A simple model describing the dynamical evolution of such structures was proposed by \citet{spruit1981} and was subsequently used by several authors \citep[see for example][]{moreno1983}.
This model has successfully reproduced Joy's law  for the tilt of bipolar regions at the solar photosphere \citep{choudhuri1987,dsilva1993,caligari1995} as well as the relationship between the tilt angle and the total flux of active regions \citep{fan1994}.
Many authors have sought to move on from this simple model and have considered the full set of compressible magnetohydrodynamic (MHD) equations (or the anelastic approximation) to study the evolution of a simplified magnetic structure mimicking a flux tube.
These models have shown that a certain amount of twist must be present inside the tube in order for it to remain coherent as it rises; such a twist corresponds to a non-zero axial current inside the flux tube.
Without twist, the flux tube is shredded apart by the vorticity generated by its own motion \citep{schussler1979,longcope1996,emonet1998,fan1998,jouve2007}.
Even more problematic than the issue of tube coherence is the interaction between the rising structures and the small-scale
convective motions.
Some studies have considered the rise of gradually twisted magnetic flux tubes through the convective zone both in Cartesian
\citep{fan2003,abbett2004} and spherical \citep{jouve2009} geometries.
Again, a given amount of twist is necessary for the flux tubes to remain coherent and rise through the convective zone.
There is observational evidence of non-potential magnetic fields in active regions \citep{seehafer1990,leka1996,pevtsov2001}, with the implication that the flux tubes are twisted before they emerge as
active regions.
While the twist is no more than one turn in general \citep{chae2005}, the corresponding Lorentz forces are implicated in chromospheric flux
eruptions \citep{torok2005}.
The origin of the twist is still unclear as there are various different explanations.
The effect of the Coriolis force \citep{fan2000} and  differential rotation \citep{devore2000} on the rising structures can only contribute a small fraction of the observed twist values \citep{holder2004}.

Other mechanisms have been suggested to explain the origin of non-potential magnetic fields.
One possibility is that the tube becomes twisted due to the effect of small-scale helical motions in the convective zone acting on the rising magnetic field, the so-called $\Sigma$-effect \citep{longcope1998}.
Another suggestion, which is closely related to the content of this paper, is that the twist arises by accretion of poloidal fields during the rise of a flux tube \citep{choudhuri2003,chatterjee2006}.
However indirect observational evidence suggests the presence of an initial twist in flux tubes at the base of the solar convective zone \citep{holder2004}, so that the twist may be produced during the formation process of the flux tube itself.

Another issue, which we don't yet fully understand, is the mere existence of localised magnetic structures similar to flux tubes below the convective zone.
What is known is that the differential rotation profile produces a strong toroidal field at the base of the solar convective zone.
We can attempt to model the appearance of isolated flux structures by looking at a uniform layer of toroidal field where the vertical structure of the magnetic layer is initially imposed; then the horizontal length scale will be set naturally by the buoyancy instability.
As a result of the magnetic pressure, the density of polytropic atmosphere in the magnetised region is reduced, assuming the temperature is not modified by the presence of the magnetic field.
Consequently, the upper interface of the magnetic slab is susceptible to Rayleigh-Taylor-type instabilities.
This paper is principally concerned with the three-dimensional evolution of these instabilities when the initial magnetic field has non-zero twist.
The two-dimensional evolution of states with no initial twist was investigated by \citet{cattaneo1988}; since no variation was permitted along the original field direction this treatment could only encompass the interchange instability, with the field lines remaining straight.
This configuration has also been considered numerically in three
dimensions by several authors \citep{matthews1995,wissink2000}, whereas alternative initial conditions (in which the temperature profile is altered by the magnetic field instead of the density) were considered by \citet{fan2001}.
It has proved very difficult to produce flux structures of significant size from this type of instability.
Whatever the scale of the original instability secondary vortex instabilities due to the down flows between the plumes disrupt the magnetic structures leaving a rather diffuse and dynamically inactive field .
Note that an alternative model has been suggested by \citet{kersale2007}, where they considered the non-linear evolution of magnetic buoyancy instabilities resulting from a smoothly stratified horizontal magnetic field.
\cite{cattaneo1990} conducted two-dimensional simulations of an initial twisted field structure.
They found that in their constrained geometry large structures could appear as result of the twist.
We will show in this paper that this effect disappears if we allow the motion to be fully three-dimensional.

The earlier simulations to examine the evolution of buoyant magnetic structures assumed that the region above the initial magnetic slab was field free.
In this case the initial potential energy contained in the initial condition is eventually transferred to kinetic energy and is finally dissipated.
Any possible effects of the magnetic field existing above the unstable slab are neglected.
Is this a reasonable simplification?

As shown by many simulations of overshooting convection in the presence of magnetic field (see for example \citet{tobias2001}), the strong toroidal field initially injected in the system or generated by a shear flow mimicking the tachocline \citep{guerrero2011} is predominantly contained in the stable radiative zone.
However, despite the effects of turbulent pumping, a non-negligible part of the field is still present in the convection zone, both due to the redistribution of large-scale magnetic fields and to local small-scale dynamo action.
If the strong toroidal field is buoyantly unstable, it will rise and interact with the small scale magnetic perturbations present in the convective zone.

The purpose of this paper is to investigate the simplest model that allows us to consider the buoyancy instability of a toroidal magnetic field in a weakly magnetised atmosphere above.
The layer of strong toroidal field models the field induced by shear in the tachocline whereas the magnetised atmosphere above represents the magnetic perturbations in the convective zone.
For this first, illustrative study, the field in the convective zone is assumed to be unidirectional but with a different direction as in the layer of strong field below.
This is of course highly idealised, but allows us to derive a model involving few free parameters.
The instability of a sheared magnetic field has already been studied (\citet{cattaneo1990,cattaneo1990b,kusano98,nozawa2005}).
In this case, the twist is uniformly distributed inside the unstable slab and the atmosphere above is field-free.
To our knowledge, this type of instability has only been considered in published papers using two-dimensional numerical simulations.
The configuration we consider later in the paper is different as the field lines are unidirectional except in a thin region where the twist and current are concentrated.
A similar setup was considered by \citet{stone2007a,stone2007b} where they look at the effect of a transverse magnetic field on the Rayleigh-Taylor instability.

The paper will proceed as follows: in section \ref{sec:model}, we describe the governing equations, the model and physical parameters and the numerical methods.
We then extend the results by \citet{cattaneo1990} and \citet{kusano98} to the three-dimensional case in section
\ref{sec:shear}.
Finally, the interchange instability in a weakly magnetised atmosphere is considered in section \ref{sec:atmo}, using both linear stability analysis and non-linear numerical simulations in three dimensions.
 
%
%
\section{Model and method \label{sec:model}}

\subsection{Model and governing equations}


We consider the evolution of a plane-parallel layer of compressible fluid, bounded above and below by two impenetrable, stress-free boundaries, a distance $d$ apart.
The upper boundary is held at fixed temperature, $T_0$ whereas a vertical temperature gradient $\theta$ is fixed at the lower boundary.
The geometry of this layer is defined by a Cartesian grid, with $x$ and $y$ corresponding to the horizontal coordinates.
The $z$-axis points vertically downwards, parallel to the constant gravitational acceleration $\bm{g}=g\hat{\bm{z}}$.
The horizontal size of the fluid domain is defined by the aspect ratios $\lambda_x$ and $\lambda_y$, so that the fluid occupies the domain $0<z<d$, $0<x<\lambda_xd$ and $0<y<\lambda_yd$.
The physical properties of the fluid, namely the specific heats $c_p$ and $c_v$, the shear viscosity $\mu$, the thermal conductivity $K$, the magnetic permeability $\mu_0$ and the magnetic diffusivity $\eta$, are assumed to be constant.
The model is identical to that used by, for example, \citet{matthews95b}, \citet{silvers2009} and \citet{favier2012} .

It is convenient to introduce dimensionless variables, so we adopt the scalings described in \citet{matthews95b} and \citet{bushby08}.
Lengths are scaled with the depth of the layer, $d$.
The temperature, $T$, and the density, $\rho$, are scaled with their values at the upper surface, $T_0$ and $\rho_0$ respectively.
The velocity, $\bm{u}$, is scaled with the isothermal sound speed, $\sqrt{R_*T_0}$, at the top of the layer, where $R_*$ is the gas constant.
We adopt the same scaling for the Alfv\'en speed, which implies that the magnetic field, $\bm{B}$, is scaled with $\sqrt{\mu_0\rho_0R_*T_0}$.
Finally, we scale time by an acoustic time scale $d/\sqrt{R_*T_0}$.

We now express the governing equations in terms of these dimensionless variables.
The equation for conservation of mass is given by
\begin{equation}
\label{eq:mass}
\frac{\partial \rho}{\partial t}+\nabla\cdot\left(\rho\bm{u}\right)=0 \ .
\end{equation}
Similarly, the dimensionless momentum equation can be written in the following form, 
\begin{multline}
\label{eq:momentum}
\frac{\partial \bm{u}}{\partial t}+\bm{u}\bm{\cdot}\bm{\nabla}\bm{u}=-\frac{1}{\rho}\bm{\nabla}P+\frac{F}{\rho}\left(\nabla\times\bm{B}\right)\times\bm{B}+\\\theta(m+1)\hat{\bm{z}}+\frac{\kappa \sigma}{\rho}\bm{\nabla}\bm{\cdot}\bm{\tau} \ ,
\end{multline}
where $P$ is the pressure given by the equation of state $P=\rho T$ and $\bm{\tau}$ is the rate of strain tensor defined by
\begin{equation}
\tau_{ij}=\frac{\partial u_i}{\partial x_j}+\frac{\partial u_j}{\partial x_i}-\frac23\delta_{ij}\frac{\partial u_k}{\partial x_k} \ .
\end{equation}
Several non-dimensional parameters appear in equation~\eqref{eq:momentum} including $\theta$, the vertical temperature gradient at the lower boundary.
Without magnetic field, it is initially equal to $\Delta T/T_0$ where $\Delta T$ is the temperature difference between the upper and lower boundaries.
$m=gd/R_*\Delta T-1$ corresponds to the polytropic index.
The dimensionless thermal diffusivity is given by $\kappa=K/d\rho_0c_p(R_*T_0)^{1/2}$ and $\sigma = \mu c_p/K$ is the
Prandtl number.
The dimensionless strength of the magnetic field is defined by $F=B_0^2/(\mu_0\rho_0R_*T_0)$, where $B_0$ is the amplitude of the imposed magnetic field.
It is related to the plasma $\beta$ by $F=2/\beta$. 
The induction equation for the magnetic field is
\begin{equation}
\label{eq:induction}
\frac{\partial \bm{B}}{\partial t}=\nabla\times\left(\bm{u}\times\bm{B}-\zeta_0\kappa\nabla\times\bm{B}\right) \ ,
\end{equation}
where $\zeta_0=\eta c_p\rho_0/K $ is the ratio of magnetic to thermal diffusivity at the top of the layer (inverse Roberts number).
The magnetic field is solenoidal so that
\begin{equation}
\nabla\cdot\bm{B}=0 \ .
\end{equation}
Finally, the heat equation is
\begin{multline}
\label{eq:heateq}
\frac{\partial T}{\partial t}+\bm{u}\bm{\cdot}\bm{\nabla}T=-\left(\gamma-1\right)T\bm{\nabla}\bm{\cdot}\bm{u}+
\frac{\kappa\gamma}{\rho}\nabla^2 T + \\ \frac{\kappa(\gamma-1)}{\rho}\left(\sigma \tau^2/2 + \zeta_0F|\bm{\nabla}
\times \bm{B}|^2\right)  \ ,
\end{multline}
where $\gamma=c_p/c_v$.

To complete the specification of the model, some boundary conditions must be imposed.
In the horizontal directions, all variables are assumed to be periodic.
As has already been described, the upper and lower boundaries are assumed to be impermeable and stress-free, which implies that $u_{x,z}=u_{y,z}=u_z=0$ at $z=0$ (the upper boundary) and $z=1$ (the lower boundary).
Having non-dimensionalised the system, the thermal boundary conditions at these surfaces correspond to fixing $T=1$ at $z=0$ and $\partial_zT=\theta$ at $z=1$.
For the magnetic field boundary conditions, we choose appropriate conditions for perfectly-conducting boundaries, which implies that $B_z = B_{x,z}=B_{y,z}=0$ at $z=0$ and $z=1$.

\subsection{Initial conditions and parameters}

In this paper, we investigate the instability of a layer of purely horizontal magnetic field.
The vertical component of the magnetic field $B_z$ is therefore initially zero and we define a vertical profile for the initial horizontal field $B_H(z,t=0)=\sqrt{B_x^2(z)+B_y^2(z)}$.
In this paper, we consider two main initial magnetic field configurations.
In section \ref{sec:shear}, the horizontal field is non-zero only in a prescribed layer $z_t<z<1$, where $z_t$ is the depth of the unstable interface and $z=1$ is the lower boundary of the numerical domain.
In section \ref{sec:atmo}, we include a weak horizontal field above the strong horizontal field in the magnetic layer to give a magnetised atmosphere throughout the domain.
This new configuration allows the magnetic tension to alter the development of the instability.
For these prescribed initial magnetic fields an equilibrium static state ($\bm{u}=\bm{0}$) can be found by solving equations \eqref{eq:momentum} and \eqref{eq:heateq} for the density and temperature profiles respectively.
The temperature gradient is slightly increased from the polytropic atmosphere due to the presence of ohmic heating, whereas the density profile is modified in order to keep the interface in pressure equilibrium (as seen later in figure \ref{fig:basic}(b)), making the layer of strong field buoyantly unstable.
This magnetised polytropic layer, coupled with a small random thermal perturbation, is an appropriate initial condition for our numerical simulations.
\begin{table}
 \centering
  \caption{Parameters values considered in this paper \label{tab:one}}
  \begin{tabular}{@{}ccc@{}}
  \hline
   Variable & Parameter & Value \\
 \hline
   $\kappa$ & Thermal diffusivity & $5\times10^{-4}$ \\
   $\sigma$ & Prandtl number & $5\times10^{-2}$ \\
   $\zeta_0$ & Inverse Roberts number & $5\times10^{-2}$ \\
   $\theta$ & Thermal stratification & $2$ \\
   $m$ & Polytropic index & $1.6$ \\
   $F$ & Dimensionless magnetic field strength & $0.1$ \\
   $\lambda_x\times\lambda_y$ & Horizontal aspect ratio & $2 \times 8$ \\
 \hline
\end{tabular}
\end{table}

There are many dimensionless parameters in this system and it is not viable to complete a systematic survey of the whole parameter space.
Throughout this paper, the polytropic index is fixed at $m=1.6$ whereas the ratio of specific heats is given by $\gamma=5/3$ (the appropriate value for a monatomic gas).
This ensures that the initial polytropic atmosphere is stable to convective motions (even with the increase in the temperature gradient due to ohmic heating) and any resulting instability is due to magnetically induced buoyancy effects.
It is also an appropriate value for the transition between the nearly-adiabatic solar convective zone and the stably-stratified radiative zone below, which is the main focus of the present model.

We fix the kinetic Prandtl number to be $\sigma=0.05$ and the inverse Roberts number to be $\zeta_0=0.05$ for the work presented in this paper.
Note that tachocline values are much lower than those that can be realised numerically.
These two values are slightly larger than the ones observed in the literature.
This is mostly justified by numerical reasons (larger values of $\sigma$ and $\zeta_0$ allow for larger time steps).
However, we have run additional simulations using smaller values for $\sigma$ and $\zeta_0$ (down to $\sigma=\zeta_0=10^{-2}$), along with cases with $\sigma<\zeta_0$ (as it should be for the tachocline), and the results are qualitatively unchanged.
Note we did not explore the possibility of the double diffusive instability \citep{silvers2009,silvers2010}, which is much more demanding numerically and will be followed up at a later point.
The thermal stratification is fixed to be $\theta=2$, which corresponds to a moderately stratified layer (the layer contains approximately three pressure scale heights).

As will be explained later, increasing the stratification is not useful in the present model.
We do not vary the dimensionless strength of the magnetic field in this paper and so $F$ is fixed to be $F=0.1$.
However we note here that we also tried to vary this parameter without any qualitative changes in the results.
The value of $F$ in the solar tachocline is expected to be much smaller than the one considered here, but this regime is challenging numerically and is therefore left for future studies.
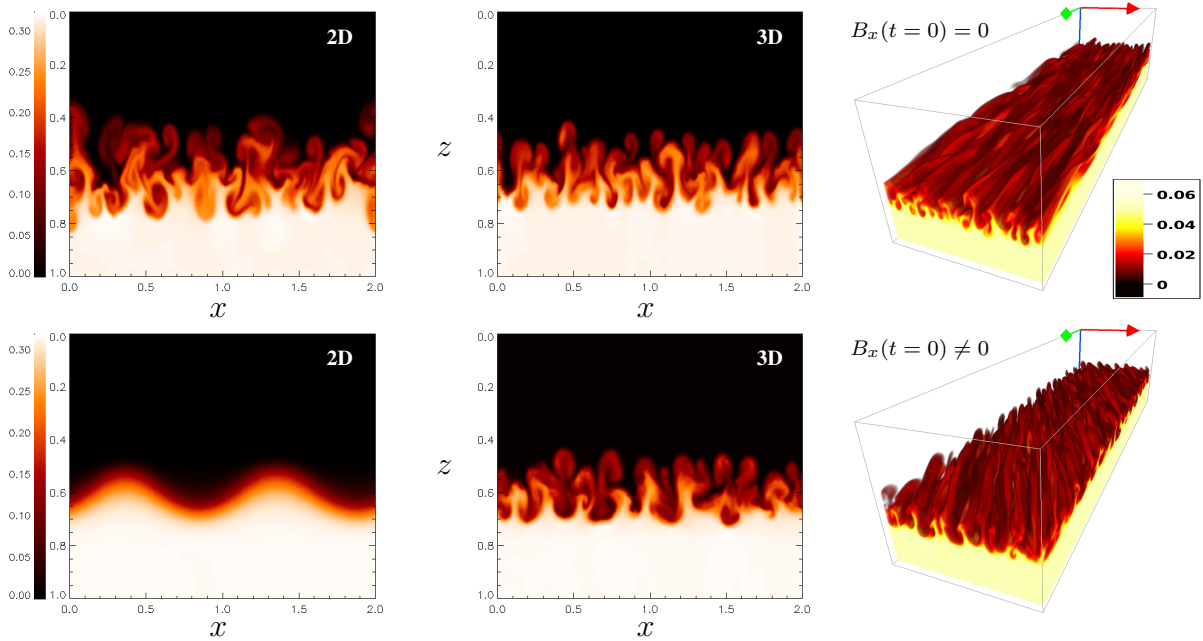
\begin{figure*}
  \begin{center}
    \begin{tabular}{ccc}
      \resizebox{52mm}{!}{\includegraphics{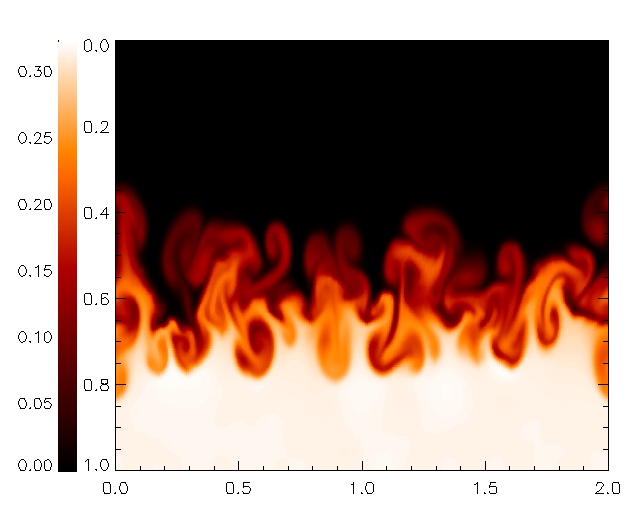}} &  
      \resizebox{52mm}{!}{\includegraphics{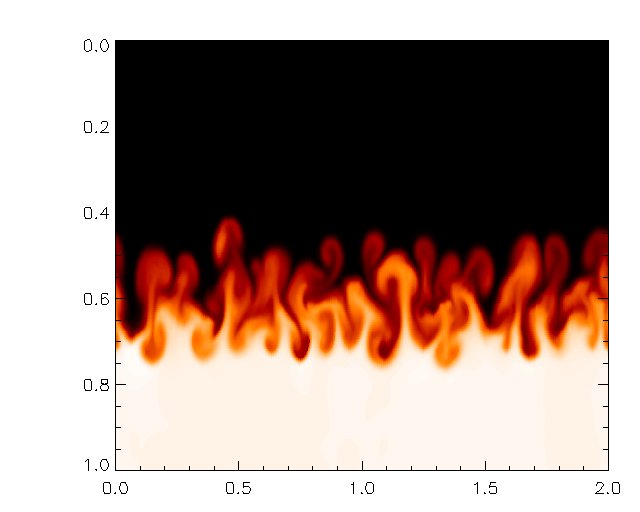}} &
      \resizebox{46mm}{!}{\includegraphics{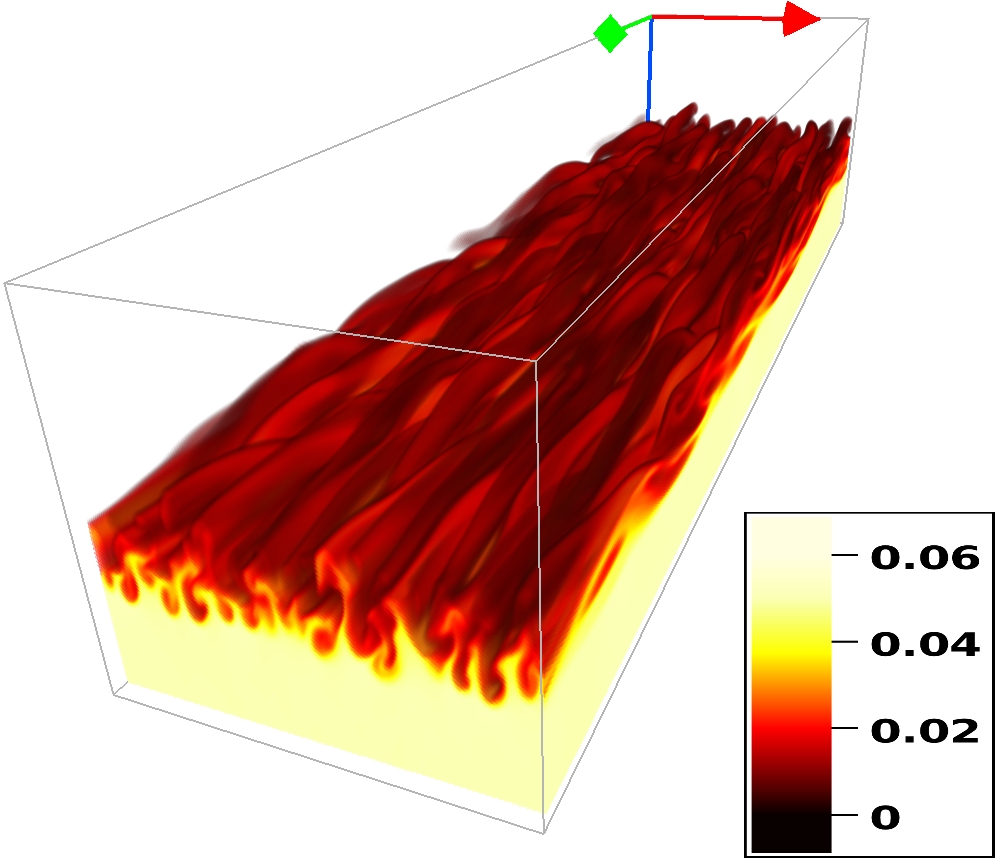}} \\
      \resizebox{52mm}{!}{\includegraphics{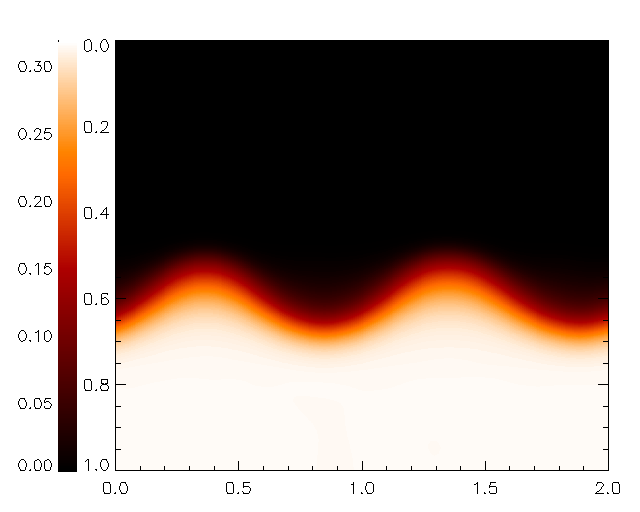}} &
      \resizebox{52mm}{!}{\includegraphics{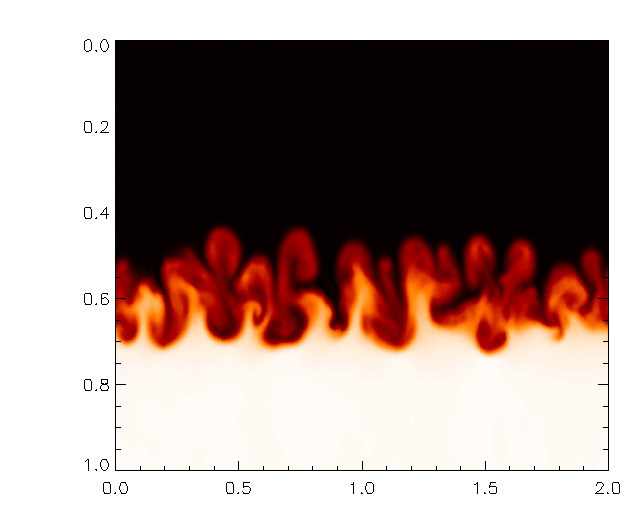}} &
      \resizebox{46mm}{!}{\includegraphics{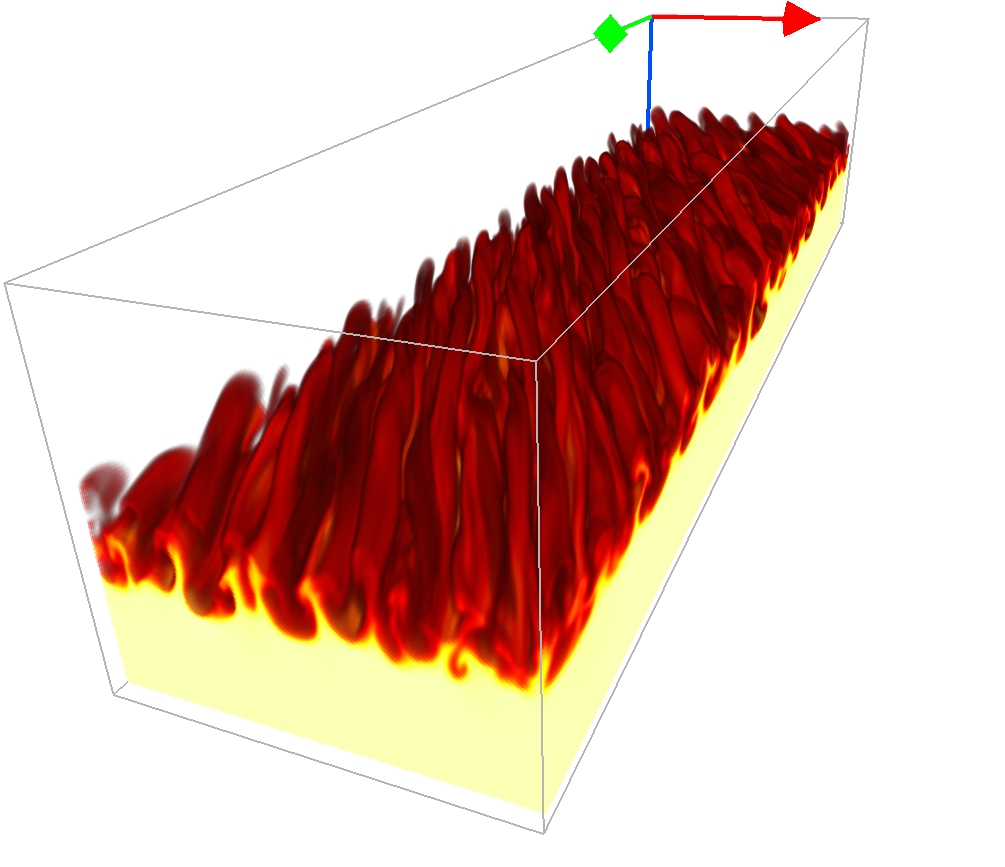}}
    \end{tabular}
    \begin{picture}(10,0)
        \put(-336,102){\textcolor{white}{\textbf{2D}}}
        \put(-175,102){\textcolor{white}{\textbf{3D}}}
        \put(-140,-15){$B_x(t=0)\neq0$}
        \put(-140,105){$B_x(t=0)=0$}
        \put(-336,-19){\textcolor{white}{\textbf{2D}}}
        \put(-175,-19){\textcolor{white}{\textbf{3D}}}
        \put(-295,60){\Large{$z$}}
        \put(-295,-60){\Large{$z$}}
        \put(-220,-1){\Large{$x$}}
        \put(-380,-1){\Large{$x$}}
        \put(-220,-121){\Large{$x$}}
        \put(-380,-121){\Large{$x$}}
    \end{picture}
    \caption{Comparison between two and three dimensional simulations. Top line: the initial field is unidirectional, directed perpendicularly to the plane. Bottom line: the initial magnetic layer is sheared. On the left column, we show contours of $B_y$ from the two-dimensional simulations, whereas the middle column corresponds to one particular $(x,z)$ plane inside the three-dimensional domain. On the right column, we show a volume rendering of the magnetic energy from the 3D simulations at the same time. All visualisations are performed at the same time $t\approx20$, except for the bottom left figure, which corresponds to $t\approx50$ (due to the fact that the growth rate of the instability is strongly reduced in this particular case see, for example, \citet{cattaneo1990}).}
    \label{fig:shear}
  \end{center}
\end{figure*}

The numerical domain is chosen to be elongated in the $y$ direction in order to accommodate for the long wavelength modulation of the secondary instability in this direction, as reported by \citet{matthews1995}.
In the transverse direction, the numerical domain is large enough to accommodate approximately $20$ most unstable wavelengths in the case without transverse magnetic field.
A summary of our choice of parameters is given in Table~\ref{tab:one}.

\subsection{Numerical method \label{sec:num}}

The given set of equations is solved using a modified version of the mixed pseudo-spectral/finite difference code originally described by \citet{matthews95b}.
Due to periodicity in the horizontal direction, horizontal derivatives are computed in Fourier space using fast Fourier transforms.
In the vertical direction, a fourth-order finite differences scheme is used and an upwind stencil is applied for the advective terms.
The time-stepping is performed by an explicit adaptive third-order Adams-Bashforth technique, with a variable time-step.
The standard resolution is $256$ grid-points in each horizontal direction and $240$ grid-points in the vertical direction.
We also consider quasi two-dimensional simulations for which the variations along the $y$ direction (along the field lines of the unstable layer) are neglected.
A poloidal-toroidal decomposition is used for the magnetic field in order to ensure that the field remains solenoidal.

A linear stability analysis of the equilibrium state is also performed in section \ref{sec:linear} to determine how the initial buoyancy instability of the strong toroidal field $B_y$ is affected by the addition of the transverse field $B_x$.
The system of equations \eqref{eq:mass}-\eqref{eq:heateq} is perturbed, linearised and the ideal gas law and the magnetic solenoidality condition are used to eliminate the pressure and one component of the fluctuating magnetic field.
All perturbations are of the form $\exp\left(\textrm{i}k_xx+\textrm{i}k_yy+st\right)$, where $s$ is the complex growth rate.
This gives rise to the a seventh order system of equations for the fluctuating variables which may be numerically approximated by
\begin{equation}
 \label{eq:linear_matrix}
 s\bm{x_{n}}=A_{n}\bm{x_{n}} \ ,
\end{equation}
where $\bm{x_{n}}$ is the vector of discretized eigenvectors and $A_{n}$ is the matrix of discretized differential operators, using a fourth-order finite difference approximation.
The boundary conditions are applied through the relevant matrix coefficients using finite difference schemes.
The resulting matrix equation is then solved for the eigenvalues $s$ and the corresponding eigenfunctions $\bm{x_{n}}$ using the Linear Algebra PACKage (LAPACK).
%
%
\section{Sheared magnetic field \label{sec:shear}}

In this section, we extend the configuration already considered by \citet{cattaneo1990} and \citet{kusano98} to the three-dimensional case.
We consider two different cases: in the first one, the initial field is unidirectional, say $B_y$, whereas in the second case, a horizontal transverse field $B_x$ is added.
For these two different cases, the horizontal magnetic field $B_H$ is the same and we compare both two-dimensional (2D) and three-dimensional (3D) simulations.
By two-dimensional simulation, we mean that all the variables are invariant in one of the horizontal direction (namely in the $y$ direction). 
The basic state is defined as follows: the initial horizontal field $B_H(z,t=0)$ only depends on $z$ and is the same in both cases,
\begin{equation}
B_H(z)=\sqrt{B_x^2+B_y^2}=\frac12\left[1.0+\textrm{erf}\left(\frac{z-z_t}{\Delta z}\right)\right] \ ,
\end{equation}
where $z_t$ is the depth corresponding the top of the unstable layer whereas $\Delta z$ corresponds to the width of the transition between the layer and the non-magnetised atmosphere above.
This choice is compatible with the boundary conditions defined in section \ref{sec:model}.
In the following, we choose $z_t=0.6$ and $\Delta z=1/60$.
The first reference case is obtained by taking $B_x=0$ so that the field is initially unidirectional, in the $y$ direction.
In the sheared case, and following \citet{cattaneo1990}, the $B_x$ is chosen to be linearly dependent on the depth $z$ according to
\begin{equation}
  B_x = \left\{
  \begin{array}{l l}
    \epsilon \left(z-z_r\right) & \quad \text{if $z_t<z<z_r$}\\
    0 & \quad \text{otherwise}\\
  \end{array} \right.
\end{equation}
(this profile is actually smoothed using error function to avoid spurious numerical effects due to discontinuities).
The depth $z_r$ at which the transverse field goes to zero is called the resonant layer by \citet{cattaneo1990}.
We choose here $z_r=0.8$.
The initial amplitude of the transverse field is derived from $\epsilon=-0.75$ (the maximum value of $B_x$ initially occurs at $z=0.6$ and is $B_x=0.15$).
Note that we tried different configurations with different resonant layers and different amplitudes for the $x$ field, but the results are qualitatively similar.
\begin{figure}
  \begin{center}
    \begin{tabular}{cc}
      \resizebox{76mm}{!}{\includegraphics{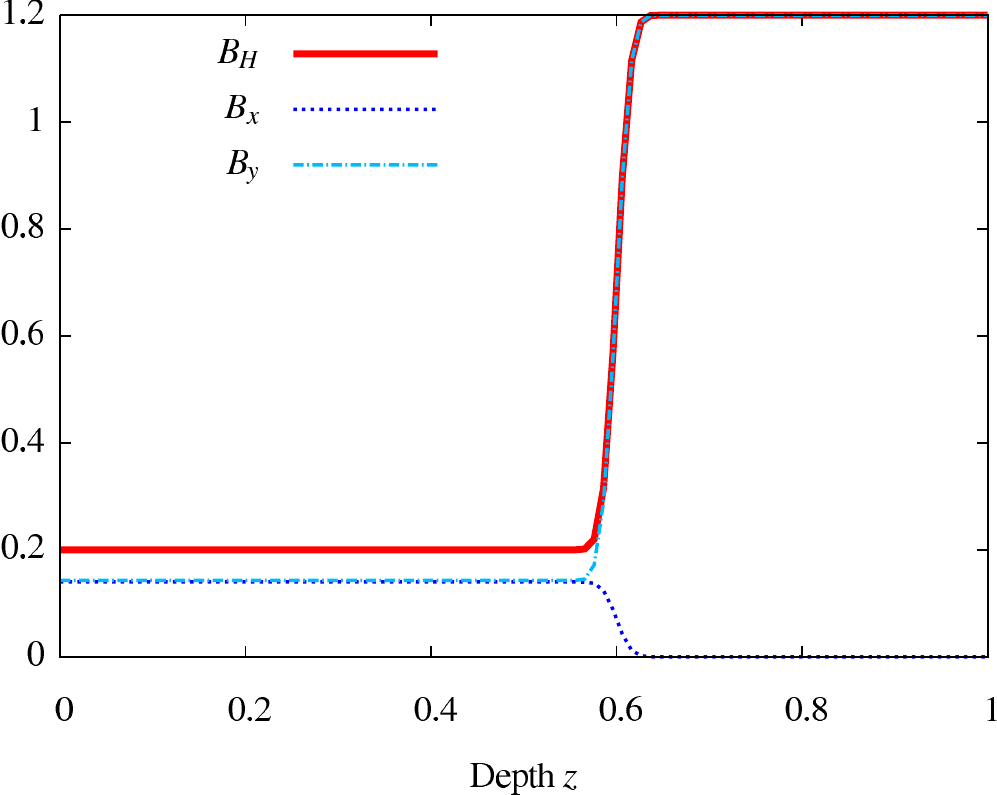}} \\
      \vspace{1mm}\\
      \resizebox{76mm}{!}{\includegraphics{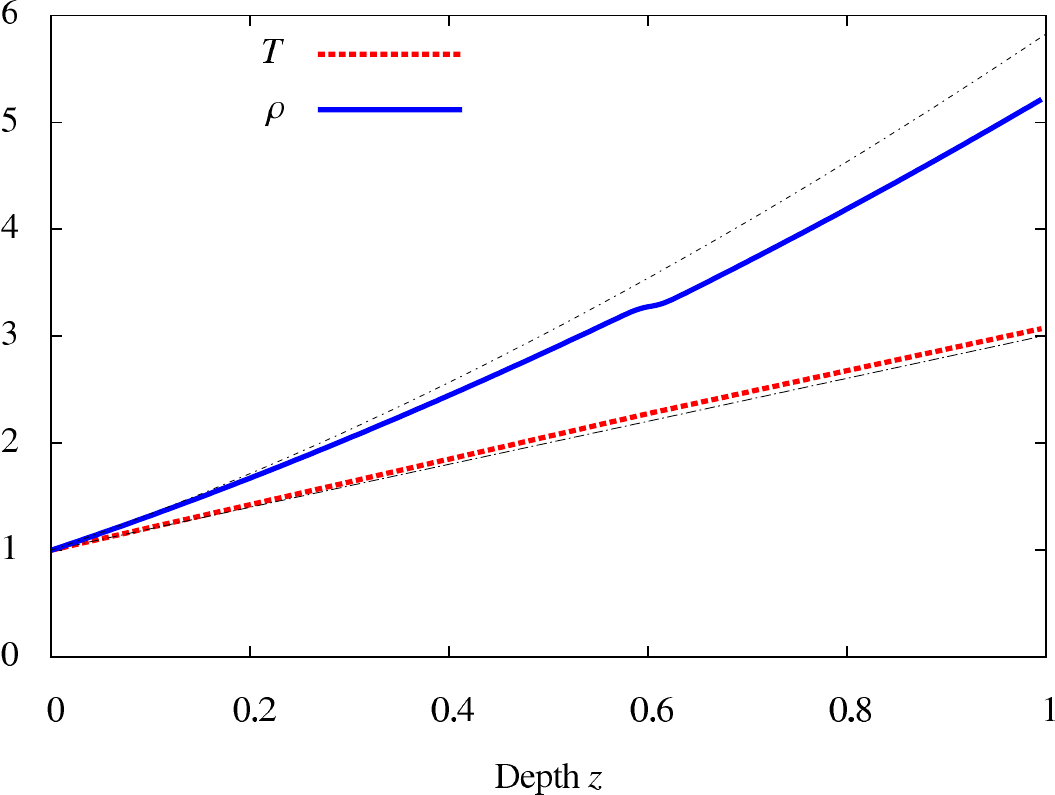}}
    \end{tabular}
    \begin{picture}(10,0)
        \put(90,220){(a)}
        \put(90,40){(b)}
    \end{picture}
    \caption{Initial condition for the horizontal components of the magnetic field, the temperature, $T$, and the density, $\rho$. In the top image, the dashed lines correspond to the vertical profile of $B_x$ and $B_y$ for the particular case $\alpha=\pi/4$. In the bottom image, the dashed lines represent the basic state of temperature and density in the absence of magnetic field. The parameter $F$ has been increased from $F=0.1$ (the value used in this paper) to $F=1$ in order to enhance the modifications of the basic field-free polytropic state.}
    \label{fig:basic}
  \end{center}
\end{figure}
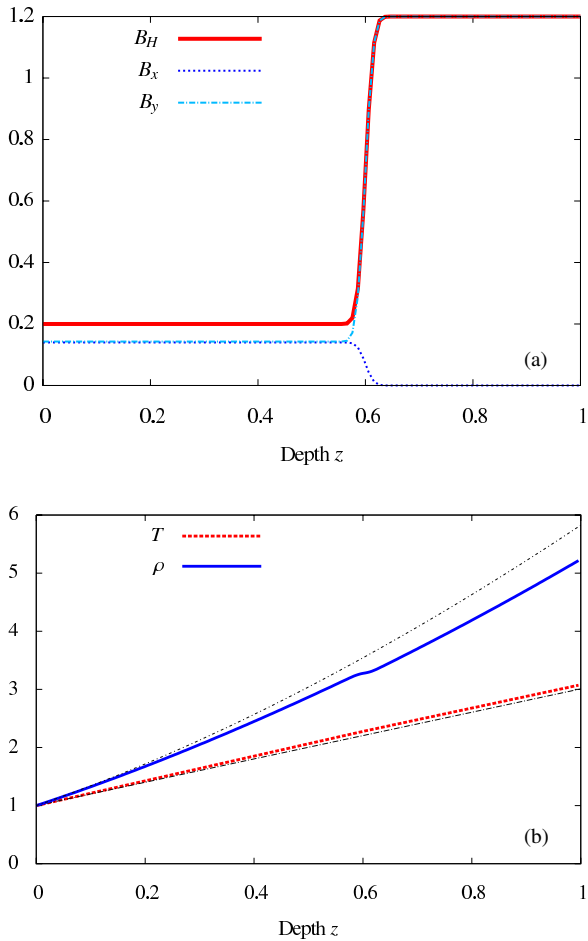

From this initial condition, equations \eqref{eq:mass}-\eqref{eq:heateq} are solved using the numerical method described in section \ref{sec:num}.
The simulations are run until the upper and lower boundary conditions have a significant effect on the generated flow.
We note here that throughout all the results, the time is presented in terms of sound crossing time units.
We present in figure \ref{fig:shear} a comparison of two-dimensional and three-dimensional simulations for the two cases with and without transverse field.
The cases with a unidirectional field $B_y$ are shown in the upper part of the figure, whereas the cases where $B_x\neq0$ initially are shown in the lower part.
It can be seen that in the unidirectional case, both 2D and 3D simulations are very similar.
The main difference is due to the lack of vortex stretching in the 2D simulation, leading to different mixing properties.
The striking similarity between 2D and 3D simulations is a confirmation of the interchange nature of the initial instability, which is purely two-dimensional.
On the other hand, in the sheared case (i.e. when $B_x \neq 0$ initially, lower line of figure \ref{fig:shear}), a drastic difference is observed between 2D and 3D simulations.
In 2D, we recover the large-scale instability already observed by several authors (see for example \citet{cattaneo1990} and \citet{kusano98}).
The small-scale interchange instability observed in the unidirectional case is altered by the tension due to the transverse field, leading to a long wavelength instability.
This strong modification is not observed in 3D, where the usual interchange instability occurs.
The similarity between the 3D simulations with and without $B_x$ field is clear.
Figure \ref{fig:shear} also shows a three-dimensional visualisation comparing the unidirectional and sheared cases using the software VAPOR\footnote{\url{http://www.vapor.ucar.edu/}} \citep{vapor1}.
It is clear that the nature of the instability is similar in both cases, apart from the change in direction of the field lines.
In 2D, due to geometrical constraints, it is not possible for the $x$ field to undergo interchange instability, whereas it is possible in 3D.
The results are similar for many different configurations (e.g. changing $z_t$, $z_r$, $\epsilon$ and the shape of the $B_x$ field from linear to quadratic).
Thus we conclude from what we have observed that imposing a transverse field inside the unstable layer is not enough to alter interchange instability in three dimensions, as was previously thought using 2D simulations.
We cannot rule out the possibility that this conclusion depends on the parameters and initial conditions considered, but we simply conclude that it might not be as easy as in the 2D case.

In the next section, we examine another initial configuration, which leads to a significant modification of the instability, while working both in two and three dimensions.
%
%
\section{Instability in a magnetised atmosphere \label{sec:atmo}}

In this section, unlike in section \ref{sec:shear} and previous studies, the atmosphere above the layer of strong field is now also magnetised.
Initially, we fix the vertical field to be zero everywhere whereas the horizontal field is only varying in the vertical direction according to
\begin{equation}
B_H=\sqrt{B_x^2+B_y^2}=0.2+0.5\left[1.0+\textrm{erf}\left(\frac{z-z_t}{\Delta z}\right)\right] \ ,
\end{equation}
where $z_t$ is the depth corresponding the top of the unstable layer and $\Delta z$ corresponds the width of the transition between the layer and the weakly magnetised atmosphere.
As in section \ref{sec:shear}, we choose $z_t=0.6$ and $\Delta z=1/60$.
In the following, the region $0<z<z_t$ is called the atmosphere whereas the region $z_t<z<1$ is called the layer.
The $x$-component of the magnetic field is derived from
\begin{equation}
\label{eq:defbx}
B_x=\frac{\epsilon}{2}\left[1.0-\textrm{erf}\left(\frac{z-z_t}{\Delta z}\right)\right] \ .
\end{equation}
The profile of horizontal field can be seen on figure \ref{fig:basic}(a), for $\epsilon=0.15$.

We define the pitch angle as being $\alpha=\textrm{atan}(B_y/B_x)$ so that $\alpha$ varies between $0$ and $\pi/2$, depending on the value of $\epsilon$. 
In the present setup, the horizontal field is six times weaker in the atmosphere than in the layer.
The actual amplitude of the field in the atmosphere is rather unimportant and we conducted additional simulations varying this parameter.
The results were qualitatively similar apart from the cases where the vertical gradients of the $x$ component of the field become strong enough to undergo interchange instability, which makes the analysis of the results difficult.
We therefore choose to focus on the simpler case where $B_x$ is large enough to have a dynamical effect on the development of the instability but small enough in order to remain stable to interchange modes during the simulations.
We would like to stress that this setup is different from the case presented in section \ref{sec:shear}, where the twist was concentrated inside the unstable layer.
Here, we consider the usual interchange instability altered by the presence of an oblique field in the atmosphere above.

\subsection{Linear analysis \label{sec:linear}}

\begin{figure}
  \begin{center}
    \begin{tabular}{c}
      \resizebox{80mm}{!}{\includegraphics{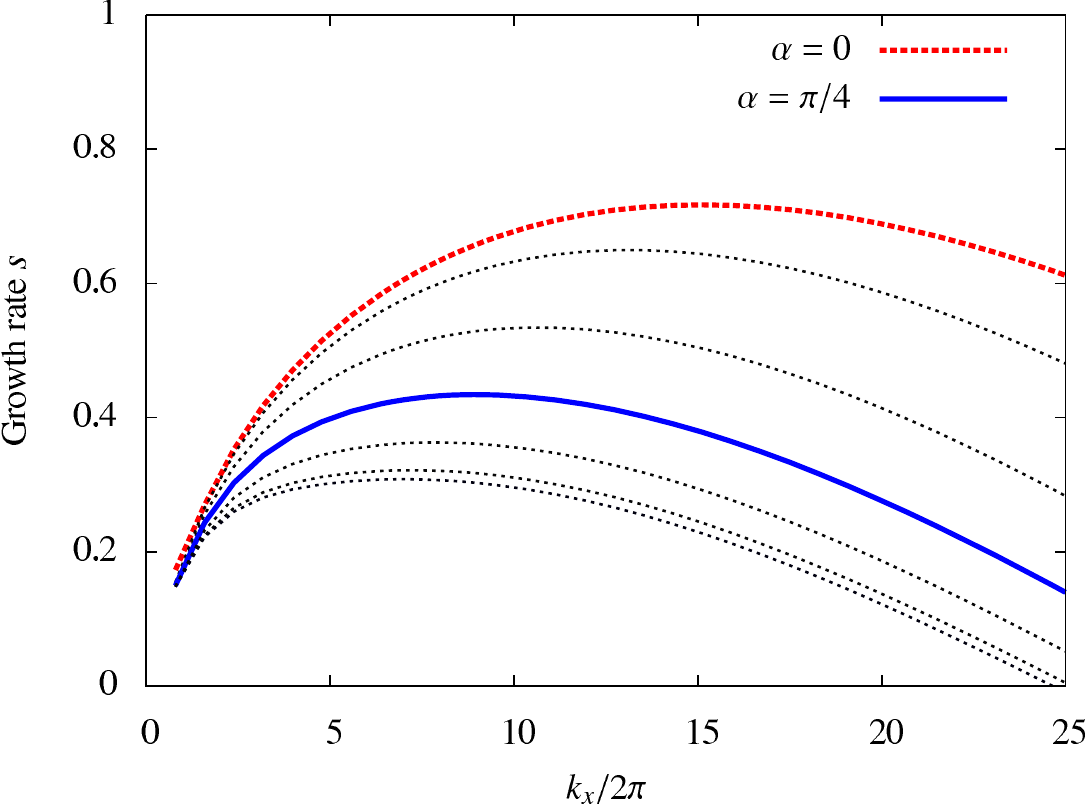}}
    \end{tabular}
    \caption{Growth rate $s$ of the instability versus horizontal wave number $k_x$ for the initial configuration presented on figure \ref{fig:basic}. The results are presented for different angles $\alpha$ from $0$ (unidirectional field) to $\pi/2$ (the field lines are perpendicular). We stress the case $\alpha=\pi/4$ as it is the largest value considered numerically in the following.}
    \label{fig:linear}
  \end{center}
\end{figure}

From the linear stability analysis we find that the most readily destabilised modes are interchange modes ($k_y=0$).
The instability is localised at the interface between the magnetic layer and the magnetised atmosphere.
Figure \ref{fig:linear} shows how the growth rate of the instability varies as a function of the horizontal wave number $k_x$ for different pitch angles $\alpha$.
As the pitch angle is increased the growth rate of the most unstable mode decreases due to the increase in magnetic tension at the interface.
This lower growth rate means that the instability will take longer to develop for larger values of the twist parameter $\alpha$.
The mode of maximum growth also decreases as the twist is increased, however the wavelength of the most unstable mode remains smaller than the depth of the domain for all values of the twist parameter $\alpha$, this will be discussed further in the following sections.
Although our configuration differs, our conclusion is qualitatively similar to the one drawn by \citet{cattaneo1990} and \citet{kusano98}, i.e. as the intensity of the transverse field is increased, both the growth rate of the instability and the transverse wave number decrease. 
%


\subsection{Non-linear 3D numerical simulations \label{sec:nonlinu3d}}

In this section, we discuss the result from non-linear simulations in 3D.
The parameter that we vary here is the initial pitch angle of the field in the atmosphere contained between $z=0$ and $z=0.6$.
We first run a reference simulation where $B_x$ is initially zero everywhere.
This corresponds to the classical interchange-type instability already considered in many papers (see, for example, \citet{matthews1995} and \citet{wissink2000}), apart from the initial presence of a weak magnetic field in the atmosphere above the layer.
Using this as a reference case, we examine three different simulations that only differ by the parameter $\epsilon$ as defined in equation \eqref{eq:defbx}.
The parameter $\epsilon$ varies from $\epsilon=0$ (unidirectional field, $\alpha=0$) to $\epsilon=0.15$ ($\alpha=\pi/4$).
The reason why we don't consider angles larger than $\pi/4$ will be explained later.
For each of these initial configurations, we solve equations \eqref{eq:mass}-\eqref{eq:heateq} using the numerical method described previously.

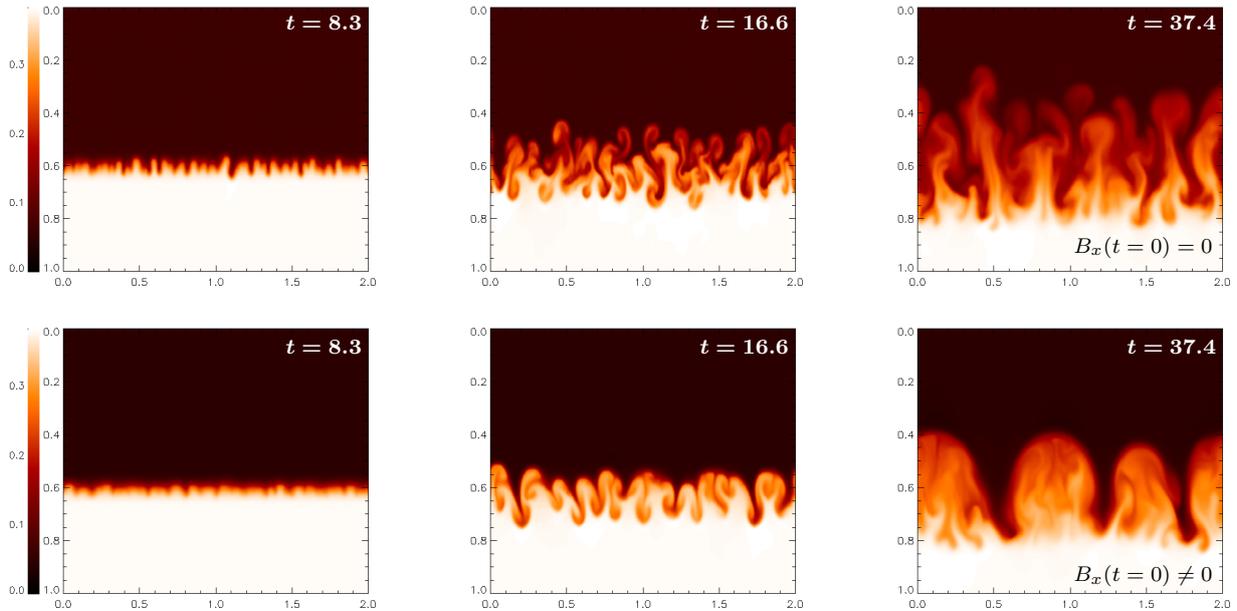
\begin{figure*}
  \begin{center}
    \begin{tabular}{ccc}
      \resizebox{52mm}{!}{\includegraphics{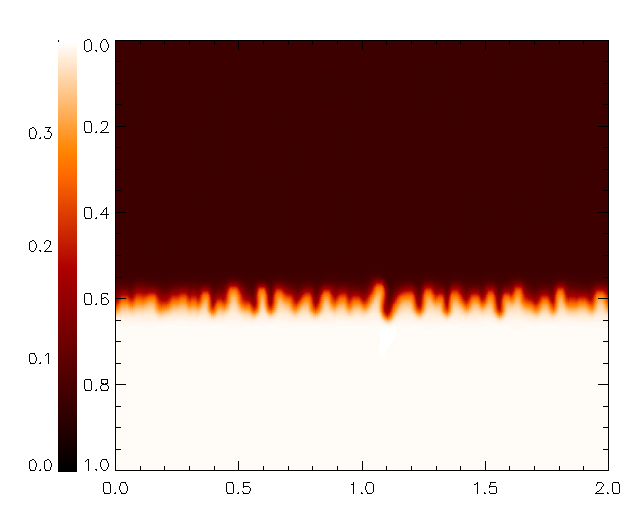}} &
      \resizebox{52mm}{!}{\includegraphics{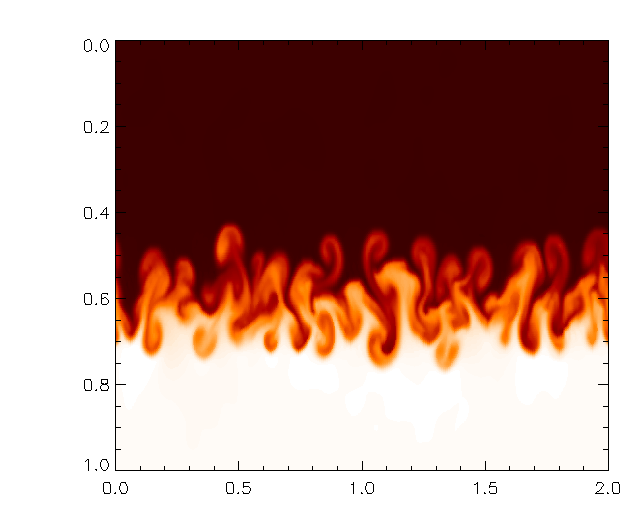}} &
      \resizebox{52mm}{!}{\includegraphics{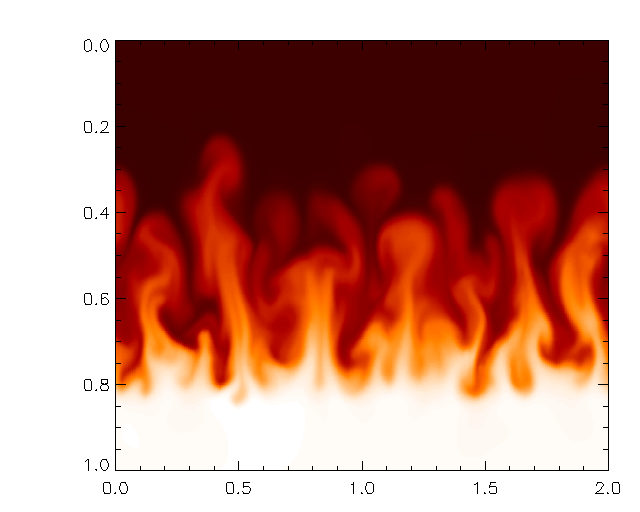}} \\
      \resizebox{52mm}{!}{\includegraphics{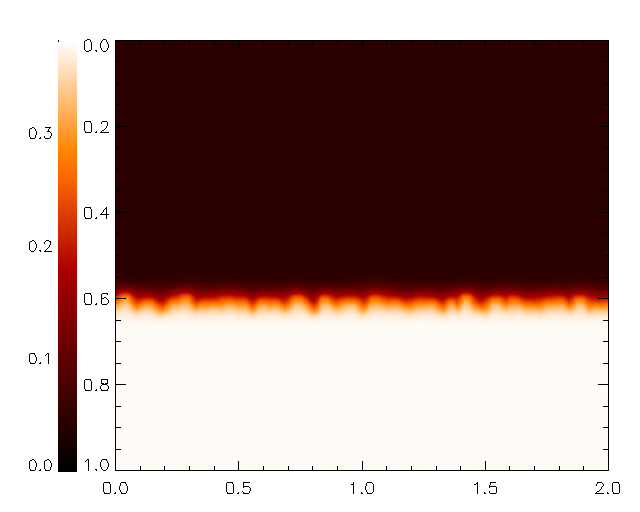}} &
      \resizebox{52mm}{!}{\includegraphics{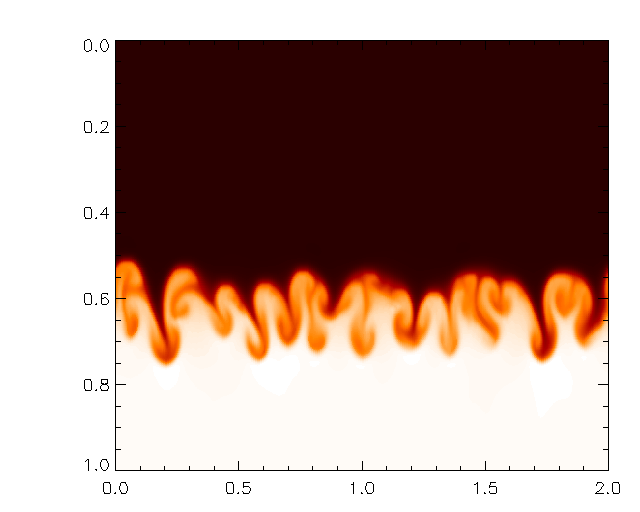}} &
      \resizebox{52mm}{!}{\includegraphics{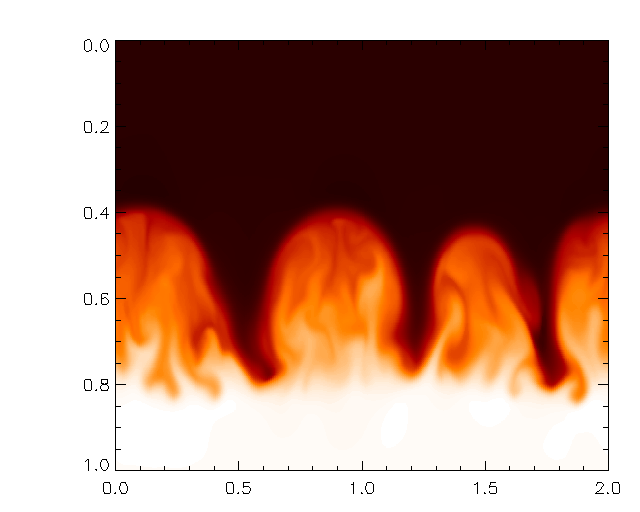}}
    \end{tabular}
    \begin{picture}(10,0)
        \put(-119,104){\textcolor{white}{$\bm{t=8.3}$}}
        \put(36,104){\textcolor{white}{$\bm{t=16.6}$}}
        \put(196,104){\textcolor{white}{$\bm{t=37.4}$}}
        \put(176,18){$B_x(t=0)\neq0$}
        \put(176,142){$B_x(t=0)=0$}
        \put(-119,226){\textcolor{white}{$\bm{t=8.3}$}}
        \put(36,226){\textcolor{white}{$\bm{t=16.6}$}}
        \put(196,226){\textcolor{white}{$\bm{t=37.4}$}}
    \end{picture}
    \caption{Evolution of the instability in the ($x$,$z$) plane perpendicular to the initial field lines in the layer. On the upper row, the magnetic field lines are unidirectional and directed perpendicularly to the plane of visualisation. On the lower row, the field lines for $z<0.6$ are orientated at $45$ degrees with respect to the field lines inside the layer. Bright colours corresponds to strong $B_y$ field whereas weaker field is dark. Time is increasing from left to right: $t=8.3$, $t=16.6$ and $t=37.4$.}
    \label{fig:0slices}
  \end{center}
\end{figure*}

We present in figure \ref{fig:0slices} vertical slices in the plane $(x,z)$ where contours of the toroidal field $B_y$ are plotted at different times.
Bright colours correspond to large values whereas dark tones correspond to low values of $B_y$.
The upper part of the figure correspond to the unidirectional field case (\textit{i.e.} $B_x=0$ initially).
As already observed in section \ref{sec:shear}, the initial instability consists of alternation of upward and downward flows.
As they increase in amplitude, a secondary instability, of Kelvin-Helmholtz type, develops in the shear region generated at the interface between the upward and downward motions.
This tends to generate horizontal vorticity, which further mixes field lines with different amplitudes of $B_y$.
The end product of this type of simulations is qualitatively the same for different parameters: the initial coherent layer is disrupted by the vortical motion generated by the secondary instability.
The resulting toroidal field is diffuse and no intense coherent magnetic structures are observed.
It is worth noting that the presence of a weak $B_y$ field in the upper part of the domain doesn't modify the nature of the instability, as can be seen comparing the top middle panel of figure \ref{fig:0slices} with the top middle panel of figure \ref{fig:shear}.
Note however that, as noted by \citet{vasil2009}, one way to damp the secondary instability is to consider very large values of the magnetic Prandtl number.
By doing so, the secondary instability is damped by strong viscosity whereas the magnetic field remains coherent.
It is however difficult to justify such a parameter regime as the magnetic Prandtl number based on molecular diffusivities is expected to be very small in the tachocline \citep{gough2007}.

Now consider a second simulation where the $x$ component of the magnetic field is initially non zero in the atmosphere above the unstable layer.
We first consider the case where the pitch angle is $\pi/4$ (as depicted in figure \ref{fig:basic}(a)) in order to illustrate the dynamical effect of the transverse field $B_x$.
Initially, as seen on the lower left panel of figure \ref{fig:0slices}, the nature of the instability is unchanged, apart from a larger characteristic horizontal length scale, as predicted by the previous linear analysis of section \ref{sec:linear}.
From the simulation, the most unstable wave number in the $x$ direction is roughly twice as small in the case $\alpha=\pi/4$ than in the reference case $\alpha=0$ (see also the magnetic energy spectra in figure \ref{fig:spect} below).
At later times ($t=16.6$, lower middle panel of figure \ref{fig:0slices}), it is apparent that the non-linear development of the instability is also modified due the magnetic tension existing at the interface between the unstable layer and the atmosphere above.
Figure \ref{fig:0slices} shows an alternation of strong and weak down flows due to the merging of small magnetic structures into larger ones.
The reduction of the growth rate as predicted by the linear analysis of section \ref{sec:linear} is also apparent, as the magnetic fluctuations are much more spread across the layer in the unidirectional case.
At the final time $t=37.4$, the difference between the two cases is very clear: the presence of the $x$ field in the atmosphere tends to generate large coherent magnetic structures in the $x$ direction containing fluctuations on smaller scales generated during the initial stages of the instability.
The toroidal field $B_y$ is also much less diffuse than in the unidirectional case, and a clear interface between field and field-free regions is observed.
At later times, the magnetic and velocity fluctuations are too close to the upper and lower boundaries, and we stop the simulations to avoid spurious confinement effects.
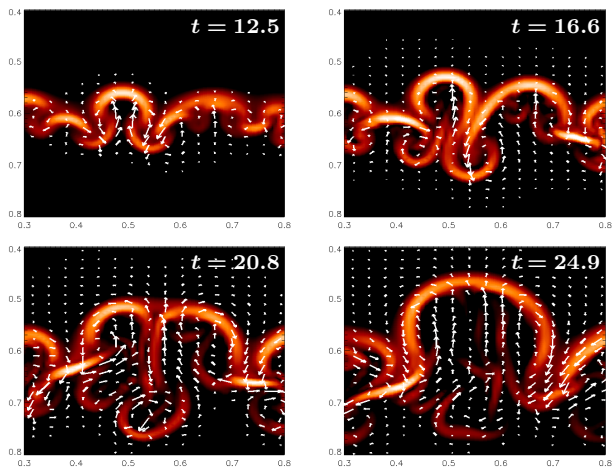
\begin{figure}
  \begin{center}
    \begin{tabular}{cccc}
      \resizebox{38mm}{!}{\includegraphics{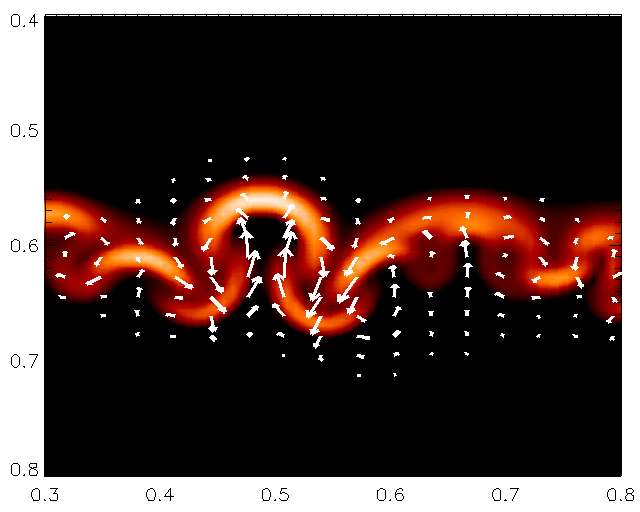}} &
      \resizebox{38mm}{!}{\includegraphics{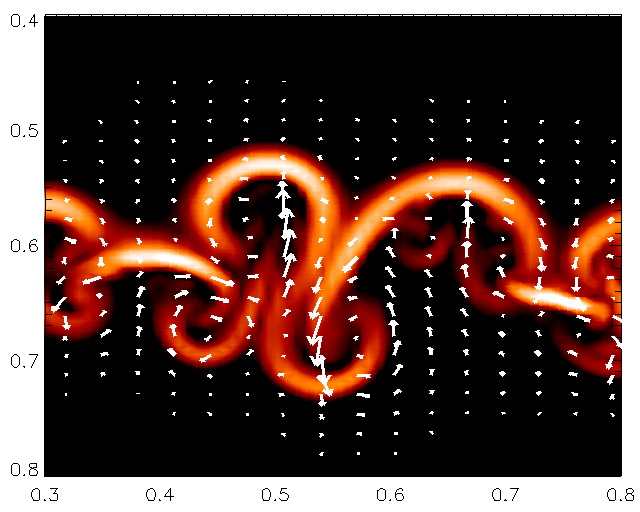}} \\
      \resizebox{38mm}{!}{\includegraphics{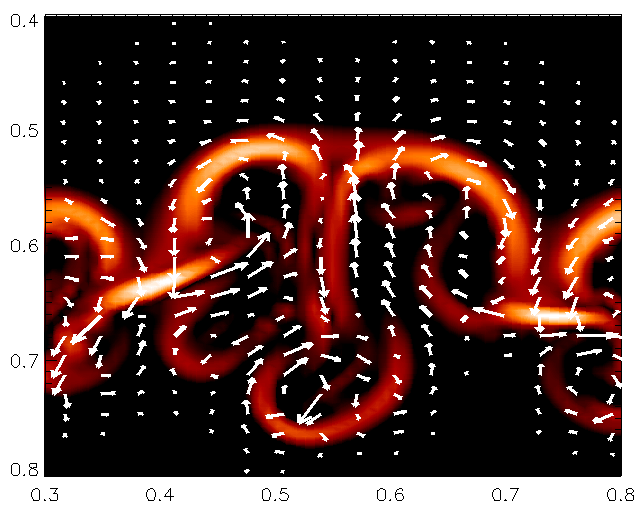}} &
      \resizebox{38mm}{!}{\includegraphics{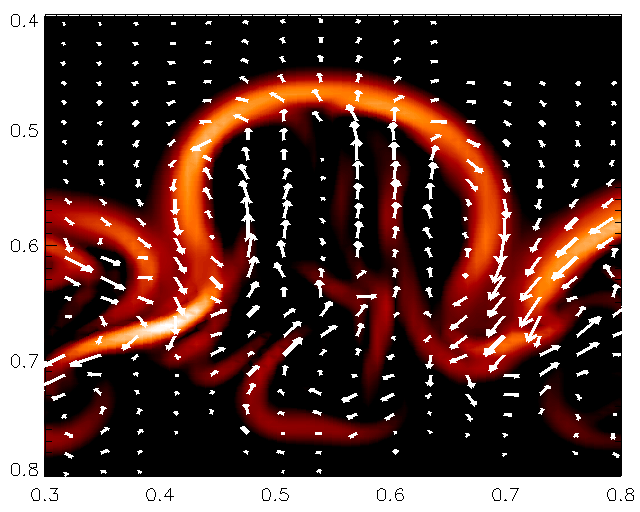}}
    \end{tabular}
    \begin{picture}(10,0)
        \put(-34,169){\textcolor{white}{$\bm{t=12.5}$}}
        \put(86,169){\textcolor{white}{$\bm{t=16.6}$}}
        \put(-34,80){\textcolor{white}{$\bm{t=20.8}$}}
        \put(86,80){\textcolor{white}{$\bm{t=24.9}$}}
    \end{picture}
    \caption{Details of the merging mechanism. The contours show the current density $\bm{j}^2$ in a small part of the $(x,z)$ plane for different times. Dark tones correspond to weak current whereas light colours correspond to strong current. The same colorscale is used for the different times. The velocity field in the plane $(x,z)$ is also shown as arrows whose length depends on the velocity amplitude.}
    \label{fig:merg}
  \end{center}
\end{figure}

It is informative to look in details at one of the merging event, where two small magnetic structures create a larger one in order to overcome the magnetic tension existing at the interface.
A time series of a subpart of the $(x,z)$ plane is presented on figure \ref{fig:merg}.
We plot both the current density $\bm{j}^2=\left(\nabla\times\bm{B}\right)^2$ as contours along with arrows representing the velocity field in this particular vertical plane.
Initially, a strong down flow in the middle of the panel is detected, dragging down the magnetic field $B_y$.
This vertical down flow must also advect the weak $B_x$ field existing in the atmosphere, therefore working against magnetic tension.
At time $t=20.8$ (third panel), it is apparent that the downward flow is now weaker and disconnected from the upper atmosphere.
The secondary instability still occurs as a vortex has been created inside the layer.
However, the vorticity is not continuously fed by the down flow.
The final stage consists of a unique magnetic structure containing small scale perturbations slowly diffusing away.
This reconnection process is clearly visible in the current density, where two rising structures are pushed together due to magnetic tension, leading to the formation of one large current sheet at the interface.  

\begin{figure*}
  \begin{center}
    \begin{tabular}{ccc}
      \resizebox{50mm}{!}{\includegraphics{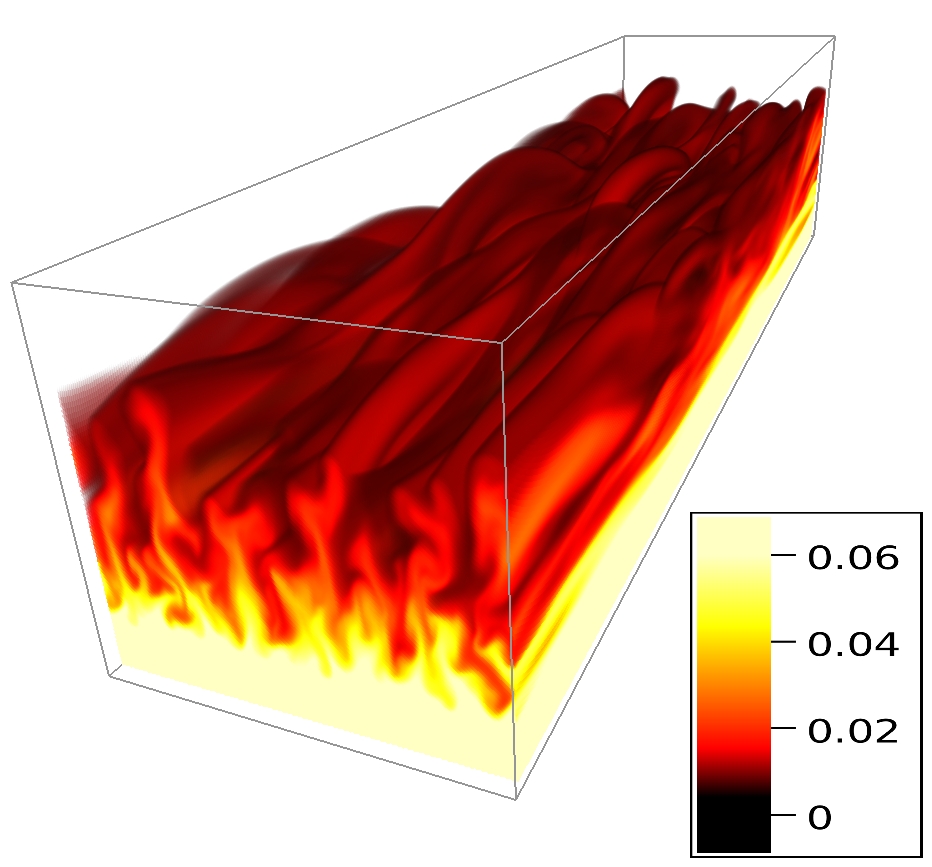}} &
      \resizebox{50mm}{!}{\includegraphics{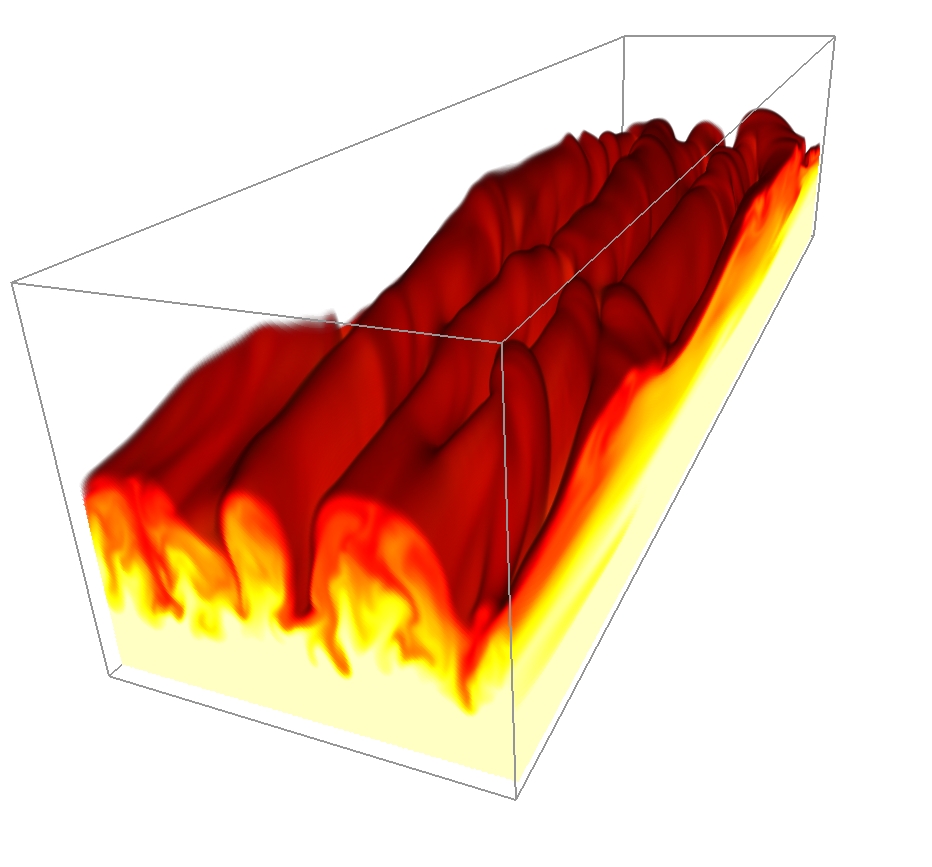}} &
      \resizebox{50mm}{!}{\includegraphics{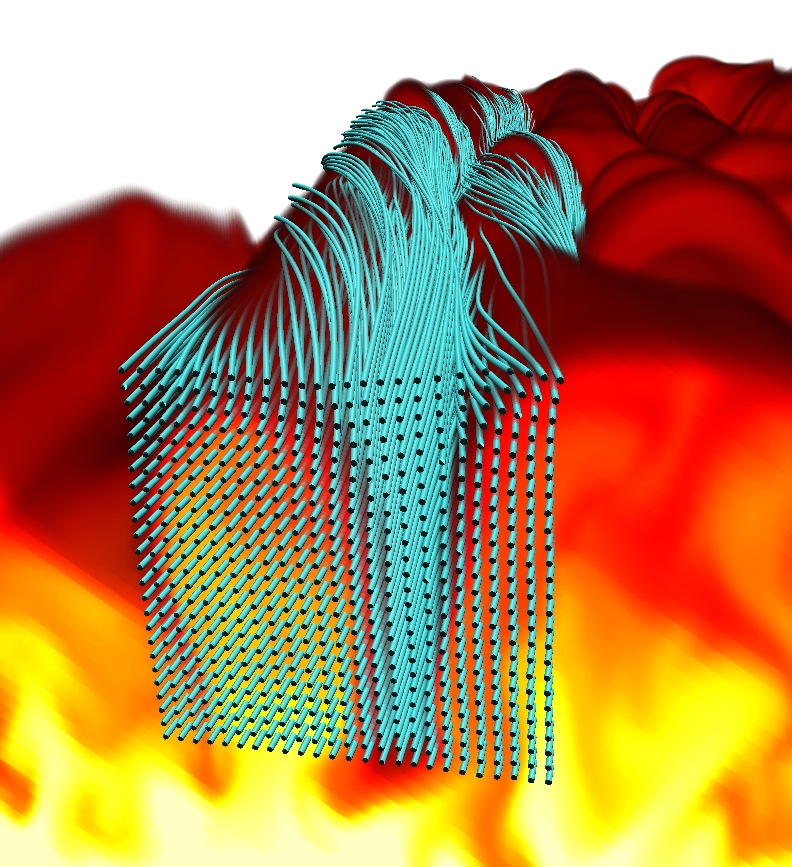}}
    \end{tabular}
    \caption{Volume rendering of the magnetic energy. The case $\alpha=0$ is presented on the left whereas the case $\alpha=\pi/4$ is on the middle. The visualisations are realised at the same time $t=24$. The figure on the right presents a subvolume of the case $\alpha=\pi/4$ where magnetic field lines have been added in order to enhance the twisted nature of the field.}
    \label{fig:vapor}
  \end{center}
\end{figure*}

Finally, and before comparing quantitatively the different simulations, we present in figure \ref{fig:vapor} a volume rendering of the magnetic energy $\bm{B}^2/2$.
Again, we compare the unidirectional case $\alpha=0$ with the case $\alpha=\pi/4$, at the same time $t=24$.
Without any initial transverse field $B_x$, the results are similar to what was already observed by \citet{matthews1995} and \citet{wissink2000}.
The initial linear instability is purely two-dimensional and corresponds to the interchange of horizontal straight field lines.
Eventually the interaction of counter-rotating vortices generated by the secondary instability leads to the arching of the magnetic structures.
Note that, although these magnetic structures are clearly observed on figure \ref{fig:vapor}, they are very weak and lack twist.
The use of isosurfaces, as in \citet{wissink2000}, can be misleading as it gives the impression the magnetic structures are coherent with a sharp interface, which is not the case.
We therefore expect that the presence of overshooting convective motions at the interface between the convective and radiative zones will inevitably disrupt these small-scale magnetic features.
In the $\alpha=\pi/4$ case however, we observe larger magnetic structures, containing a significant amount of twist.
Note that these larger structures are far from being well-defined localised flux tubes \citep{cattaneo2006}.
We however stress that the fact that our polytropic atmosphere is stably-stratified and the magnetic tension resulting from the presence of the oblique $B_x$ field tends to act against the rise of these structures.
These limiting effects are the results of our over-simplified model, but more refined approaches could eventually lead to the formation of localised flux tubes from the break-up of a magnetic layer.
The last panel on the right in figure \ref{fig:vapor} shows a close-up with several magnetic field lines in blue.
As the strong $B_y$ field rises, twist accumulates at the apex of the magnetic structure, damping its disruption by vortical motions and increasing its longevity against eventual convective motions.
This is reminiscent of the scenario suggested by \citet{choudhuri2003}, where the twist is generated by the rise of strong toroidal fields inside the convective zone containing a poloidal field.
Note however that \citet{choudhuri2003} considered already formed flux tubes and not a uniform layer of magnetic field.
\begin{figure*}
  \begin{center}
    \begin{tabular}{ccc}
      \resizebox{52mm}{!}{\includegraphics{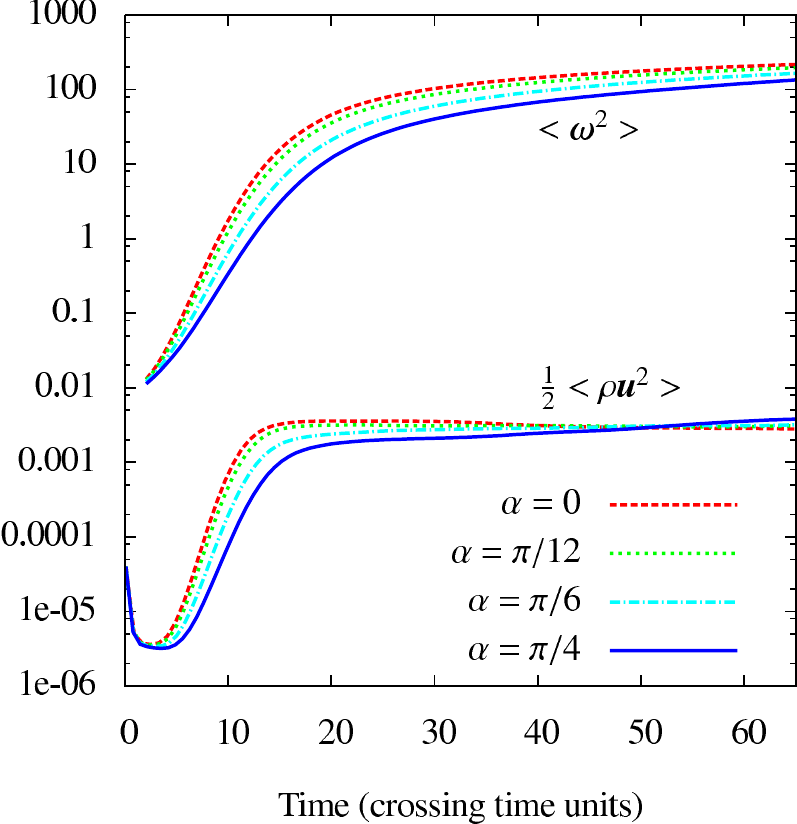}} &
      \resizebox{50mm}{!}{\includegraphics{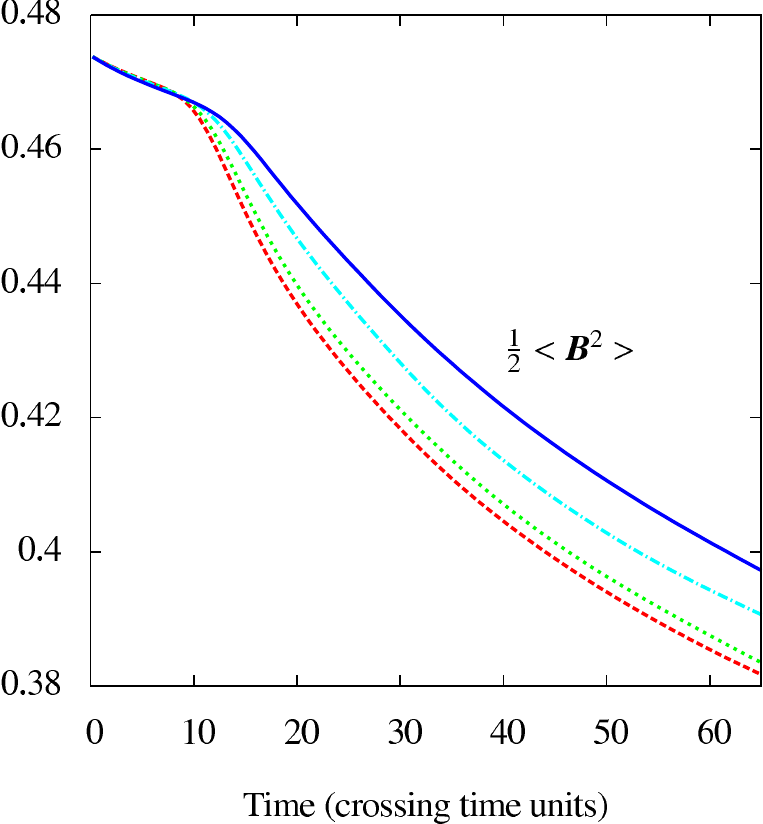}} &
      \resizebox{50mm}{!}{\includegraphics{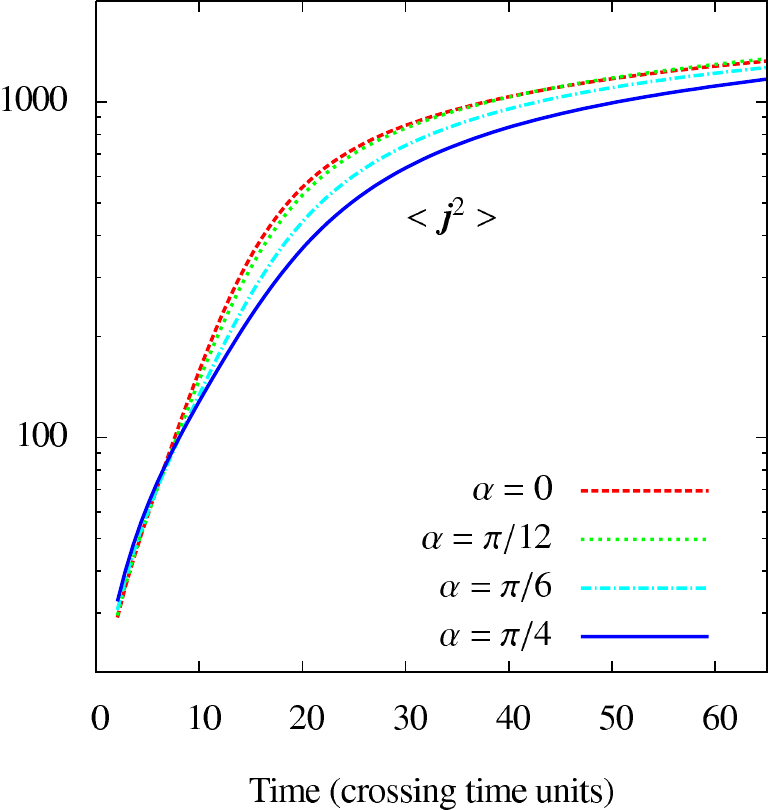}}
    \end{tabular}
    \caption{Evolution of the total kinetic energy, enstrophy, magnetic energy and current density versus time for four different angles from $\alpha=0$ to $\alpha=\pi/4$.}
    \label{fig:vort}
  \end{center}
\end{figure*}

We now discuss more quantitatively the effect of the transverse field $B_x$ on the development of the instability.
Figure \ref{fig:vort} shows the evolution with time of several global quantities.
In the following, the brackets $\left<.\right>$ denote a spatial average over all coordinates.
In particular, we define the total kinetic energy $\left<\rho\bm{u}^2\right>/2$ and the total magnetic energy $\left<\bm{B}^2\right>/2$.
As previously, we compare the results from simulations without initial $B_x$ field ($\alpha=0$) and with various pitch angles.
After the decay of the initial perturbation, the kinetic energy grows exponentially for a few sound crossing times, as seen on the left panel of figure \ref{fig:vort}.
Note that, as predicted by the linear analysis of section \ref{sec:linear}, the growth rate decreases when $\alpha$ increases.

After approximately ten sound crossing times, the instability saturates.
The evolution after this stage strongly depends on $\alpha$.
For the case $\alpha=0$, the kinetic energy decreases with time.
In this case, most of the initial potential energy contained in the initial condition is released during the linear phase, leading to a decaying non-linear regime.
For $\alpha=\pi/4$, much less kinetic energy is initially released due to the magnetic tension at the interface, leading to a growth of the kinetic energy with time up to the end of the simulation.
Of course, as no external source of energy is included in the model, we expect the kinetic energy to ultimately decay in all cases.
One of the important steps of the magnetic buoyancy instability is the generation of strong horizontal vortices by the secondary Kelvin-Helmholtz instability.
It is clear that the $x$ field, if strong enough, will tend to damp this generation of vorticity.
In order to create strong down and up flows, it is now necessary to work against the tension of the weak magnetic field $B_x$.
The evolution of the total enstrophy $\left<\bm{\omega}^2\right>$ in the numerical domain is plotted versus time on figure \ref{fig:vort}, where $\bm{\omega}=\nabla\times\bm{u}$.
In all cases, we observe a rapid increase of the enstrophy with time.
As the instability saturates, the growth rate of the enstrophy decreases.
As for the kinetic energy, we expect the enstrophy to ultimately decay.
The main effect of the transverse field is to reduce the production of enstrophy by the secondary instability, so that the enstrophy is always smaller in the cases where $\alpha \neq 0$ than in the case where $\alpha=0$.
This leads to a reduction of the viscous dissipation, therefore reducing the rate at which the kinetic energy is dissipated.

The two others panels in figure \ref{fig:vort} show the evolution with time of the total magnetic energy and current density $\left<\bm{j}^2\right>$, where $\bm{j}=\nabla\times\bm{B}$.
In all cases, the magnetic energy decays whereas the current density increases with time.
As $\alpha$ increases, the magnetic energy decreases less rapidly.
This is due to the damping of the vortical motions that reduces the mixing of the magnetic field lines and the creation of dissipative currents.
Note that the current is initially slightly larger when $\alpha \neq 0$, as expected from the initial conditions.

Following \citet{stone2007a}, it is useful to quantify the amount of mixing due to the instability.
Instead of defining the mixing in terms of density, as for the Rayleigh-Taylor instability, we define the fraction of strong horizontal field as
\begin{equation}
f_s=\frac{B_H-B_H(z=0)}{B_H(z=1)-B_H(z=0)}
\end{equation}
where $B_H=\sqrt{B_x^2+B_y^2}$.
$B_H(z=0)$ corresponds to the initial amplitude of the horizontal field at the upper boundary, and can be considered as the weak field.
$B_H(z=1)$ is the initial horizontal field at the lower boundary and is therefore the strong field.
The fraction of weak field is simply $f_w=1-f_s$. 
The amount of mixing in the system can be estimated by the following quantity
\begin{equation}
\label{eq:mix}
\Theta(z)=4\left<f_sf_w\right> \ ,
\end{equation}
which is equal to $0$ when the averaged horizontal field is equal to its initial value at the vertical boundaries, and $1$ if the field is perfectly mixed and is equal to the average between these two values.
\begin{figure}
  \begin{center}
    \begin{tabular}{c}
      \resizebox{75mm}{!}{\includegraphics{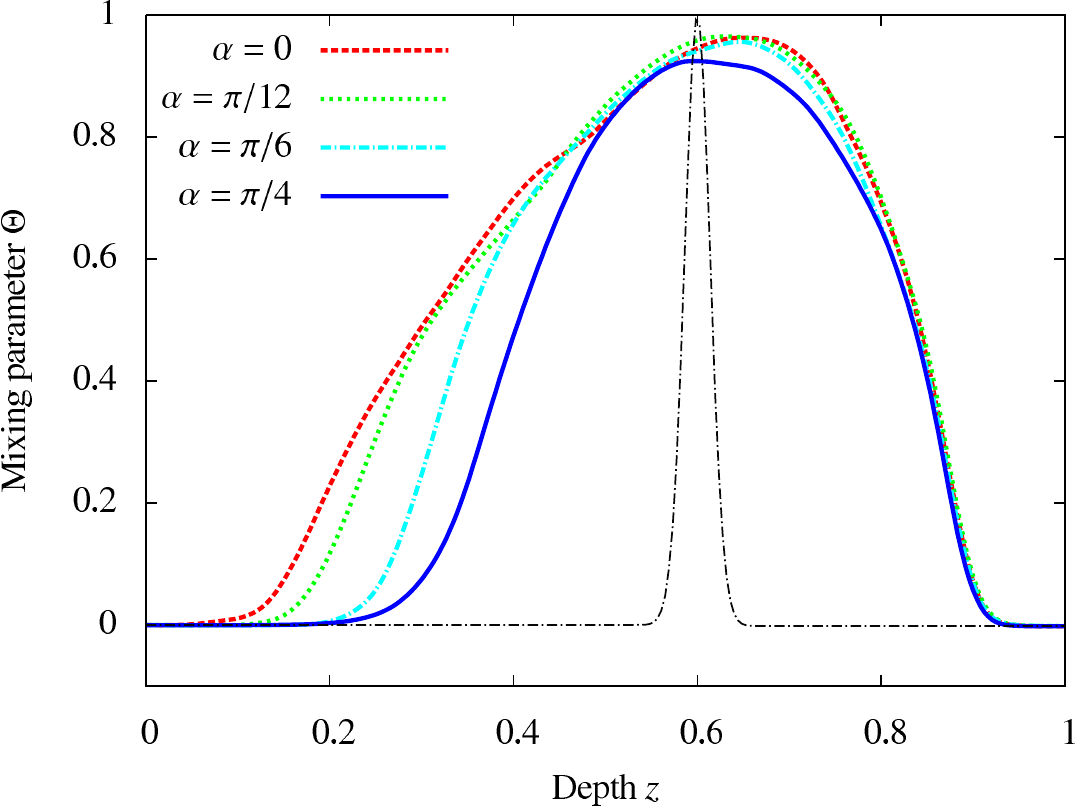}} 
    \end{tabular}
    \caption{Mixing parameter $\Theta(z)$ as defined by equation \eqref{eq:mix} versus depth at time $t=30$. The thin black line corresponds to the initial profile at $t=0$.}
    \label{fig:mix}
  \end{center}
\end{figure}

Figure \ref{fig:mix} shows the mixing parameter $\Theta$ as a function of $z$ for the four cases considered at $t=30$.
The thin black line corresponds to the value of $\Theta$ derived from the initial condition, which is nearly the same in all four cases.
As time increases, the two fronts of mixing propagate vertically.
It is clear that due to the magnetic tension at the interface, the overall mixing of the upper front is reduced.
The vertical extent of the mixing region is reduced as $\alpha$ increases.
This result can be explained by the reduced growth rate on the initial linear instability (see figure \ref{fig:linear}) in the presence of a transverse field and by the damping of the secondary instability leading to weaker vortical motions (see figure \ref{fig:vort}). 
In the unidirectional case, the potential energy is released to counter the stable stratification, leading to the rise of the toroidal magnetic field.
When $\alpha\neq0$, the magnetic tension also acts against the instability.
This is a clear limitation of our simple model, and will be further discussed in section \ref{sec:conclusions}. 
Note finally that figure \ref{fig:mix} also shows that the lower front is roughly unchanged by the presence of transverse field.
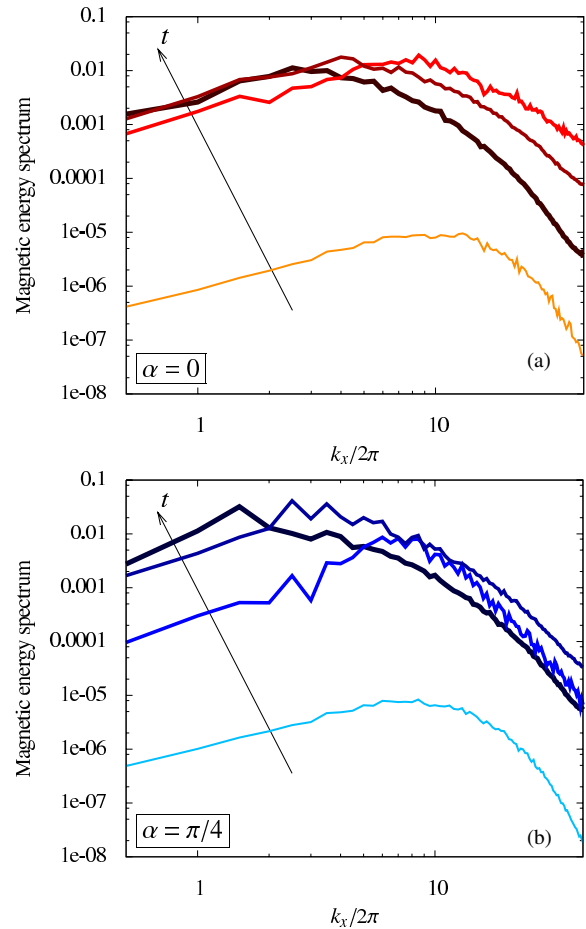
\begin{figure}
  \begin{center}
    \begin{tabular}{c}
      \resizebox{75mm}{!}{\includegraphics{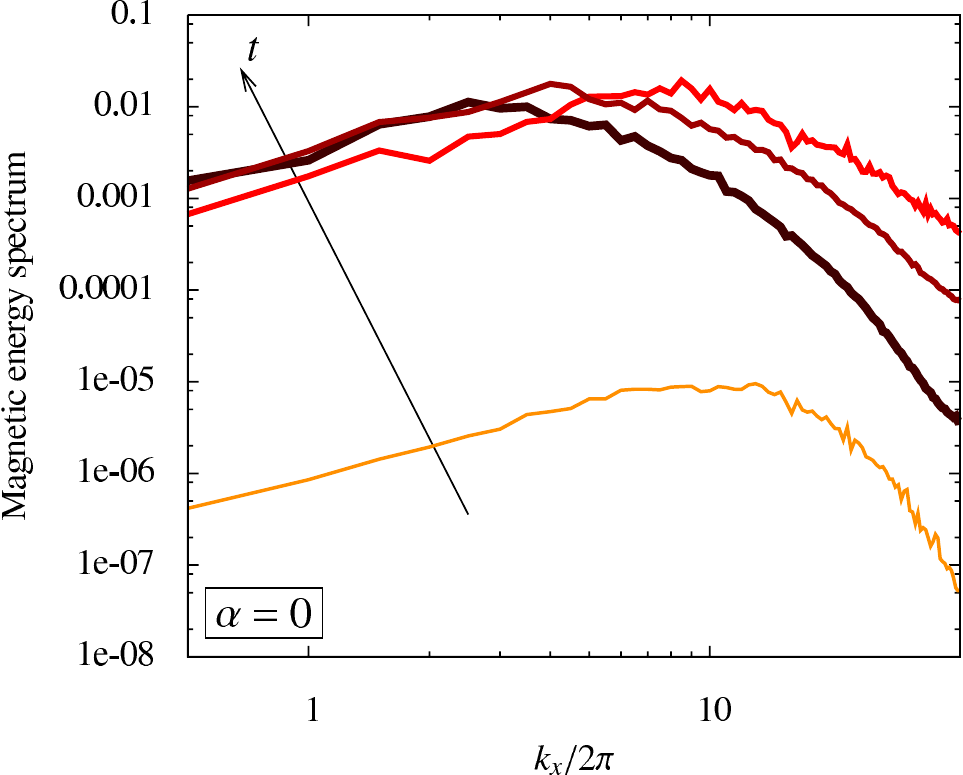}} \\
      \resizebox{75mm}{!}{\includegraphics{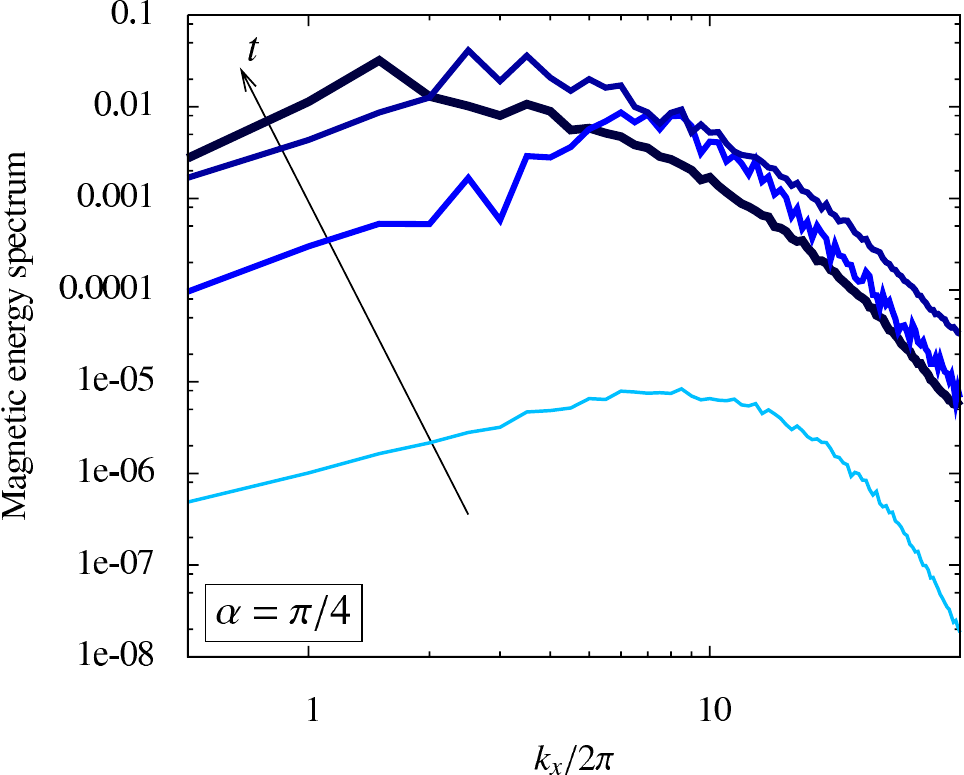}}
    \end{tabular}
    \begin{picture}(10,0)
        \put(90,215){(a)}
        \put(90,35){(b)}
    \end{picture}
    \caption{Magnetic energy spectrum $E_M(k_x)$ as defined by equation \eqref{eq:spect} at four different times $t=2$ (light tones), $t=12.5$, $t=24$ and $t=32$ (dark tones). (a) $\alpha=0$ and (b) $\alpha=\pi/4$. The results are obtained by spatially averaging over $y$ and $z$.}
    \label{fig:spect}
  \end{center}
\end{figure}

The generation of large-scale coherent magnetic structures is clearly visible in the magnetic energy spectra defined by
\begin{equation}
\label{eq:spect}
E_M(k_x)=\frac12\sum_{z}\sum_{y} \hat{\bm{B}}(k_x,y,z)\cdot\hat{\bm{B}}^*(k_x,y,z)
\end{equation}
where $\hat{\bm{B}}(k_x,y,z)$ is the one-dimensional Fourier transform in the $x$ direction of $\bm{B}(x,y,z)$, whereas the star denotes the complex conjugation.
We focus here on the magnetic energy spectra in the direction perpendicular to the initial strong field (the results in the $y$ direction are rather similar between the different cases).

Figure \ref{fig:spect} shows the magnetic energy spectrum for $\alpha=0$ and $\alpha=\pi/4$, at four different times spanning from the initial linear instability ($t=2$) to the late stage of the nonlinear phase ($t=32$).
Initially, the most unstable horizontal wave number corresponds to the linear results already discussed in section \ref{sec:linear}.
The most unstable wave number is roughly $k_x/2\pi\approx14$ for $\alpha=0$ and $k_x/2\pi\approx8$ for $\alpha=\pi/4$, which can be compared to the results presented in figure \ref{fig:linear}.
The time evolution of the magnetic energy spectra can be described in two phases.
First, the small-scale instability grows exponentially from $t\approx2$ to $t\approx12$.
The effect of the transverse field is to reduce the initial linear growth rate, as already discussed in figures \ref{fig:linear} and \ref{fig:vort}.
From $t\approx12$, the evolution of the instability is constrained by the mixing due to the vortical motions.
Here, the effect of the transverse field is to reduce the direct cascade of magnetic energy toward small scales, as we observe steeper energy spectra for $\alpha=\pi/4$ than for $\alpha=0$, especially at early times.
Finally, during the late stage of the instability, we observe a decay of the magnetic energy at small scales whereas the magnetic energy at large scales remains constant for $\alpha=0$.
In contrast, the magnetic energy is still growing at large scales for $\alpha=\pi/4$.
The effect of the transverse field is indeed clearly visible on the small wave numbers region $k_x/2\pi<3$.
For $\alpha=\pi/4$, a clear peak is observed at large scales, which corresponds to the coherent magnetic structures already observed on figures \ref{fig:0slices} and \ref{fig:vapor}.
This is also consistent with the larger values of the global magnetic energy observed on figure \ref{fig:vort} when $\alpha\neq0$.
Without transverse field, the magnetic energy peaks at larger horizontal wave numbers.
Note that this difference cannot be attributed to linear effects, as the most unstable wave numbers as predicted by the linear analysis are much larger (see figure \ref{fig:linear}).
During the non-linear phase, the small-scale features (\textit{i.e.} $k_x/2\pi>4$) are roughly independent of $\alpha$, and we observe a continuity of magnetic scales up to the resistive scale.
There is no clear scaling in our case since the diffusivities do not allow for a fully-developed turbulent state.
\begin{figure}
  \begin{center}
    \begin{tabular}{cc}
      \resizebox{78mm}{!}{\includegraphics{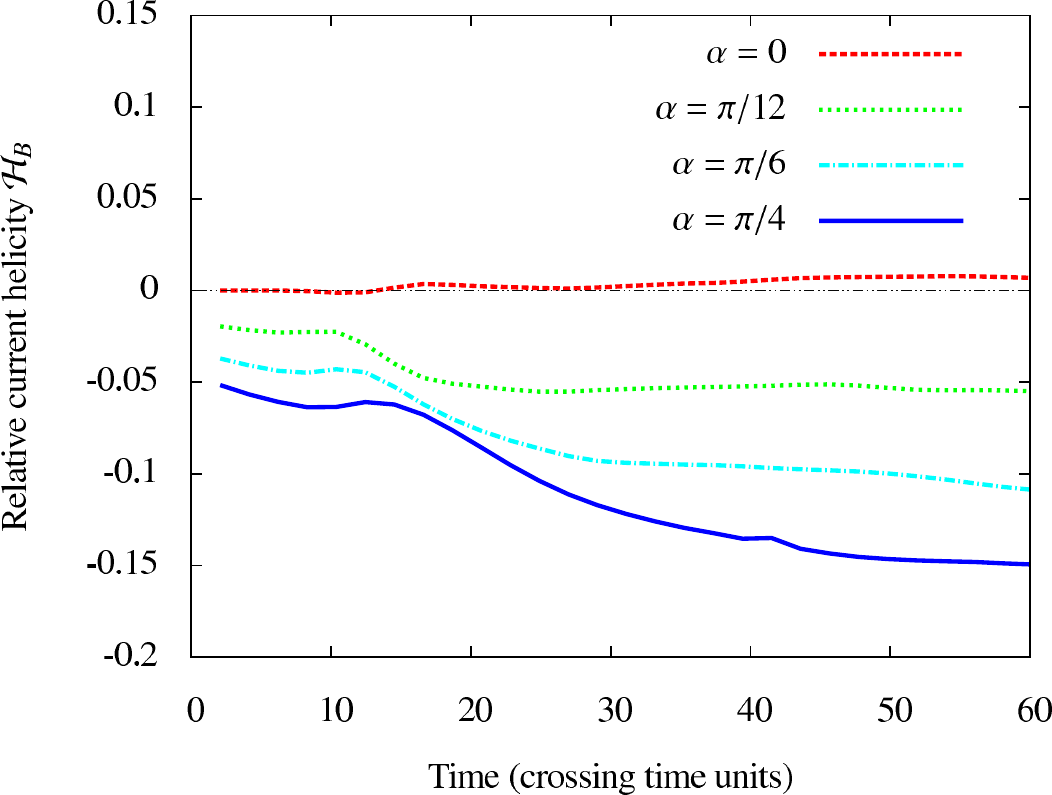}} 
    \end{tabular}
    \caption{Relative current helicity, $\mathcal{H}_B$, as defined by equation \eqref{eq:relhel} versus time.}
    \label{fig:jhel}
  \end{center}
\end{figure}

Finally, the amount of twist in the magnetic field lines, as observed in figure \ref{fig:vapor}, can be estimated looking at the current helicity, defined by $\left<\bm{j}\cdot\bm{B}\right>$.
Since both magnetic energy and current density are time-dependent (see figure \ref{fig:vort}), we consider the relative current helicity defined by
\begin{equation}
\label{eq:relhel}
\mathcal{H}_B=\frac{\left<\bm{j}\cdot\bm{B}\right>}{\sqrt{\left<\bm{j}^2\right>}\sqrt{\left<\bm{B}^2\right>}} \ .
\end{equation}
Note that given our initial configuration, the amount of current helicity depends on $\alpha$.
We show on figure \ref{fig:jhel} the evolution with time of $\mathcal{H}_B$ for different values of $\alpha$.
The effect of the initial condition is clearly apparent, as $\mathcal{H}_B=0$ when $\alpha=0$, whereas $\mathcal{H}_B$ initially increases with $\alpha$.
Without initial transverse field, the current helicity remains statistically negligible, although we observe a very weak positive correlation, which might be due to the interaction between the vorticity and the rising magnetic field.
However, as soon as $\alpha\neq 0$, we observe an increase of the relative current helicity from its small initial value.
This is particularly visible in the $\alpha=\pi/4$ case.
After a transient stage where $\mathcal{H}_B$ increases and then decreases, we observe a monotonic increase of the relative current helicity up to the end of the simulation.
Note that $\mathcal{H}_B$ continues to increase even after the saturation of the initial linear phase, which roughly corresponds to $t\approx20$ (see figure \ref{fig:vort}).
Again, since the atmosphere is stably stratified, we expect the magnetic field to ultimately stop rising and the current helicity to ultimately decay.
If the entropy gradient was nearly negligible, as expected in the convective zone, the magnetic structures observed in our simple model could continue to rise unimpeded, eventually becoming kink-unstable \citep{fan1999}.
The increase in the relative current helicity observed here is an indication that the twist observed in active regions could be due to the interaction between a rising toroidal structure and the weaker field inside the convective zone, as suggested by \citet{choudhuri2003}.
Of course, due to numerical constraints, the large diffusivities considered in this paper do not allow for the formation of isolated structures.

%
%
%
%
\section{Discussion\label{sec:conclusions}}

In this work, we sought to examine the instability of a layer of horizontal magnetic field in a polytropic atmosphere, where the direction of the field lines is depth-dependent.
In section \ref{sec:shear} we examined the instability of a layer of horizontal magnetic field in a polytropic atmosphere, where the direction of the field lines in the layer varied with depth.
We showed that the initial idea of building large-scale coherent magnetic structures from the buoyancy instability of a sheared magnetic layer, as studied by \cite{cattaneo1990} and \cite{kusano98}, seemed limited to two-dimensional geometry.
The same configuration drastically changed in three dimensions, where the interchange instability was able to occur in an oblique direction.
We stress that it could be possible to find a configuration of this type that strongly modifies the nature of the magnetic buoyancy instability, however this was not the aim of the paper.

In section \ref{sec:atmo} we introduced another initial configuration, where the twist was concentrated at the interface between a strong layer of horizontal magnetic field and a weakly magnetised atmosphere above.
The weakly magnetised atmosphere was justified by the presence of convection, which is expected to contain non-negligible magnetic fluctuations, both due to the redistribution of large-scale magnetic fields, and due to local small-scale dynamo action.
The presence of a strong unidirectional field was justified by the presence of radial differential rotation in the tachocline.
We presented a linear stability analysis and numerical simulations of the magnetic buoyancy instability in sections \ref{sec:linear} and \ref{sec:nonlinu3d}.
We have shown that the initial interchange instability is only weakly modified by the presence of the transverse magnetic field.
During the non-linear phase of the instability, the small-scale magnetic structures merged together to overcome magnetic tension, leading to the formation of magnetic structures on scales larger than those of the initial linear instability.
These structures were shown to be more coherent than those in the unidirectional case, since the production of vorticity by the secondary instability was reduced.
In addition we found a significant amount of twist was generated due to the rise of the toroidal field in the weakly magnetised atmosphere.
This mechanism could provide the initial twist necessary for the magnetic structures to rise coherently through the convective zone.

It is however important to stress the limitations of the present model.
In order to limit the number of parameters, we have considered two unidirectional fields with a current sheet at the interface.
One could argue that the magnetic field in the convective zone is interacting with small-scale motions and is certainly not unidirectional as assumed here.
Although it remains to be verified using a more refined model, we expect that locally the toroidal field will behave similarly in the presence of small-scale magnetic fluctuations.

There are two stabilising effects in the present model, the stably stratified atmosphere (which is necessary to suppress convective motions) and the magnetic tension generated by the twisted atmosphere.
The tension becomes important as the atmosphere is compressed by the rising structures, and the toroidal magnetic field will eventually stop rising.
We have reached this regime in some of our simulations where the pitch angle is equal or larger that $\pi/4$.
The stabilising effect of the tension is clearly overestimated by the present model, as the tension exists everywhere at the interface between the rising layer and the atmosphere.
In a more realistic situation, the convective zone would be filled with small-scales magnetic structures.
Locally these structures could have a similar effect to the global tension considered in our simulations, although we didn't check it numerically here.
We plan in a forthcoming paper to look at a configuration where the global magnetic tension is removed at some depth, leading the rise and formation of local coherent twisted magnetic structures.
Another possibility would be to extend the work of \citet{fan2003} and \citet{jouve2009}, assuming that the convective zone is filled with magnetic perturbations.
This would confirm whether the generation of twist, via the interaction of a rising flux tube and small-scale magnetic perturbations, is sufficient for the flux tube to remain coherent as it rises through the convective zone.
Such simulations are underway in spherical geometry using the ASH code \citep{pinto2012}.

We conclude by arguing that the merging mechanism illustrated by this simple model could have implications on the formation of large-scale magnetic structures from a deep-seated toroidal field.
The twist observed in active regions could therefore be attributed to a progressive accumulation of local twist as the toroidal field rises through the magnetised convective zone.

{\bf Acknowledgements} This work has been financially supported by STFC. CPU time was provided by the HPC resources of CALMIP under the allocation 2012-P1118 and by the UKMHD supercomputing facility located in Warwick. 
%
%
\bibliographystyle{mn2e}
\bibliography{biblio}
\label{lastpage}

\end{document}